\documentclass[aps,pre,showpacs,groupedaddress,twocolumn]{revtex4-1}

\usepackage{amsmath}
\usepackage{amssymb}
\usepackage{graphicx}
\usepackage{color}
\usepackage{bm}
\usepackage{subfigure}
\usepackage{multirow}

\begin{document}
\title{Stochastic lattice model of synaptic membrane protein domains}

\author{Yiwei Li}

\author{Osman Kahraman}

\author{Christoph A. Haselwandter}

\affiliation{Department of Physics \& Astronomy and Molecular and Computational Biology Program, Department of Biological Sciences, University of Southern California, Los Angeles, CA 90089, USA}

\begin{abstract}
Neurotransmitter receptor molecules, concentrated in synaptic membrane domains along with scaffolds and other kinds of proteins, are crucial for signal transmission across chemical synapses. In common with other membrane protein domains, synaptic domains are characterized by low protein copy numbers and protein crowding, with rapid stochastic turnover of individual molecules.
We study here in detail a stochastic lattice model of the receptor-scaffold
reaction-diffusion dynamics at synaptic domains that was found previously to capture, at the mean-field level, the self-assembly, stability, and characteristic size of synaptic domains observed in experiments. We show that our stochastic lattice model yields quantitative agreement with mean-field models of nonlinear diffusion in crowded membranes. Through a combination of analytic and numerical solutions of the master equation governing the reaction dynamics at synaptic domains, together with kinetic Monte Carlo simulations, we find substantial discrepancies between mean-field and stochastic models for the reaction dynamics at synaptic domains. Based on the reaction and diffusion properties of synaptic receptors and scaffolds suggested by previous experiments and mean-field calculations, we show that the stochastic reaction-diffusion dynamics of synaptic receptors and scaffolds provide a simple physical mechanism for collective fluctuations in synaptic domains, the molecular turnover observed at synaptic domains, key features of the observed single-molecule trajectories, and spatial heterogeneity in the effective rates at which receptors and scaffolds are recycled at the cell membrane. Our work sheds light on the physical mechanisms
and principles linking the collective properties of membrane protein domains to the stochastic dynamics that rule their molecular~components.

\end{abstract}
\maketitle

\section{Introduction}
\label{secIntro}

The stability and plasticity of synapses are thought to play central roles in memory formation and learning \cite{McAllister2007,Citri2008}. In particular,
neurotransmitter receptor molecules, concentrated in postsynaptic membrane domains along with scaffolds and other kinds of proteins \cite{Ziv2014,Salvatico2015}, are crucial for signal transmission across chemical synapses \cite{Citri2008,Legendre2001,Specht2008,Tyagarajan2014}. The strength of the transmitted signal depends on the number of receptors localized in synaptic domains \cite{Citri2008,Specht2008}, and regulation of the receptor number in synaptic domains provides one mechanism for postsynaptic plasticity \cite{Carroll2001,Shepherd2007,Kneussel2014}. With the advent of \textit{in vivo} superresolution light microscopy \cite{Hell1994,Choquet2003,Huang2010,Kusumi2014}, the multiscale properties of synaptic domains---from the stochastic diffusion
trajectories of individual synaptic receptors to the overall size and stability
of synaptic domains---can now be studied quantitatively \cite{Choquet2013,Ziv2014,Kneussel2014,Salvatico2015}. A central discovery \cite{Choquet2013,Ziv2014,Kneussel2014,Salvatico2015} here is that synaptic receptors \cite{Choquet2003,Triller2008,Triller2005}, as well as their associated scaffolds \cite{Okabe1999,Gray2006,Calamai2009,Ziv2014},
turn over rapidly, with individual molecules entering and leaving synaptic
domains on typical time scales as short as seconds. In contrast, synaptic domains of a well-defined characteristic size can persist over months or even longer periods of time \cite{Trachtenberg2002,Grutzendler2002}, which
may \cite{Citri2008,Carroll2001,Shepherd2007,Specht2008,Choquet2013,Ziv2014,Kneussel2014,Salvatico2015} constitute part of the cellular basis for memory formation and learning.

To explain the stability and characteristic size of synaptic domains in the face of rapid molecular turnover, a number of phenomenological models for membrane domain formation \cite{Sekimoto2009,Holcman2006,Earnshaw2006,Burlakov2012,Czondor2012,Shouval2005}---based
on, for instance, balancing receptor fluxes into and out of synaptic domains
or spatially varying effective reaction and diffusion rates---have been proposed. Through an interplay between quantitative experiments on minimal model systems lacking most of the synaptic machinery (for instance, single transfected fibroblast cells) and theoretical modeling it has been demonstrated 
\cite{Kirsch1995,Meier2000,Meier2001,Borgdorff2002,Dahan2003,Hanus2006,Ehrensperger2007,Calamai2009,Haselwandter2011,Haselwandter2015} that the reaction and diffusion properties of receptors and their associated scaffolds at the cell membrane are sufficient for the self-assembly of stable synaptic receptor-scaffold domains of the characteristic size observed in neurons. In particular, the presence of a presynaptic terminal is not essential for the self-assembly of stable synaptic receptor-scaffold domains. The reaction-diffusion model of synaptic domains \cite{Haselwandter2011,Haselwandter2015}
describing these experiments explains, based on a reaction-diffusion (Turing) instability \cite{Turing1952} of the mean-field equations governing receptor-scaffold reaction-diffusion dynamics, how interactions between receptors and their associated scaffolds, together with the diffusion properties of each molecule species at the cell membrane, are sufficient for the spontaneous formation, stability, and characteristic size of synaptic domains. For molecular reaction and diffusion rates consistent with experimental measurements on synaptic receptors and scaffolds 
\cite{Kirsch1995,Meier2000,Meier2001,Borgdorff2002,Dahan2003,Hanus2006,Ehrensperger2007,Calamai2009,Haselwandter2011,Haselwandter2015},
the reaction-diffusion model yields \cite{Haselwandter2011,Haselwandter2015},
starting from random initial conditions, the self-assembly, stability, and characteristic size of synaptic domains observed in neurons. Conversely,
it has been shown \cite{Calamai2009,Haselwandter2011,Haselwandter2015} that self-assembly of synaptic domains can be prevented in both experiment and theory through, for instance, selective modification of the reaction properties of scaffolds.

In common with other membrane protein domains \cite{Lang2010,Simons2010,Rao2014,Recouvreux2016},
synaptic domains are characterized \cite{Choquet2003,Specht2008,Triller2005,Triller2008,Ribrault2011,Choquet2013,Ziv2014,Salvatico2015}
by low protein copy numbers ($\approx10$--1000) and protein crowding. Previous
work
\cite{Butler2009,Butler2011,Cao2014,Lugo2008,Erban2009,Hecht2010,Erban2007,Cohen2005,Wylie2006,Samoilov2006,McKane2004,McKane2005,Gillespie1976,Muratov2005,Muratov2007,Lizana2008,Fanelli2010,Fanelli2013}
suggests that the coupling between molecular noise and the nonlinear reaction-diffusion dynamics induced by protein crowding can lead to a rich interplay between fluctuations and deterministic dynamics at synaptic domains. Indeed, experiments \cite{Ribrault2011,Choquet2013} and theoretical modeling \cite{Shouval2005,Holcman2006,Sekimoto2009,Burlakov2012,Czondor2012} indicate that synaptic domains undergo collective fluctuations that may affect synaptic signaling. In a previous paper \cite{Kahraman2016} we demonstrated
that the stochastic lattice model associated with the mean-field reaction-diffusion
dynamics at synaptic domains \cite{Haselwandter2011,Haselwandter2015} yields,
for molecular reaction and diffusion rates consistent with experimental measurements on synaptic receptors and scaffolds 
\cite{Kirsch1995,Meier2000,Meier2001,Borgdorff2002,Dahan2003,Hanus2006,Ehrensperger2007,Calamai2009,Haselwandter2011,Haselwandter2015},
emergence of synaptic domains in the presence of rapid stochastic turnover of individual molecules, provides a quantitative link between the molecular noise inherent in reaction-diffusion processes and collective fluctuations in synaptic domains, and allows prediction of the stochastic dynamics of individual synaptic receptors and scaffolds at the cell membrane. In the present article we build on this previous work \cite{Kahraman2016} to provide a detailed discussion of the stochastic lattice model of receptor-scaffold
reaction-diffusion dynamics at synaptic domains \cite{Haselwandter2011,Haselwandter2015,Kahraman2016}
and its relation to the corresponding mean-field model. We show \cite{Kahraman2016} that molecular noise can yield substantial deviations from mean-field results
for the receptor-scaffold reaction-diffusion dynamics at synaptic domains, and that stochastic lattice models can be employed successfully to provide quantitative insights into the single-molecule and collective dynamics
of membrane protein domains \cite{Lang2010,Simons2010,Rao2014,Recouvreux2016,Choquet2003}.

This article is organized as follows. We first summarize, in Sec.~\ref{secModel},
the stochastic lattice
model of receptor-scaffold reaction-diffusion processes at synaptic domains \cite{Kahraman2016,Haselwandter2015}, which is defined mathematically by a suitable master equation (ME), and its relation to the corresponding mean-field model \cite{Haselwandter2015,Haselwandter2011}
formulated in accordance with the standard formalism of chemical dynamics
\cite{Butler2011,Cao2014,Lugo2008,Erban2009,Butler2009,Hecht2010,Erban2007,Cohen2005,Wylie2006,Samoilov2006,McKane2004,McKane2005,Gillespie1976,Gillespie1977,Gillespie2013,Epstein1998,Walgraef1997,Cross2009,Cross1993,Murray2002,Meinhardt1982,Maini2001}.
We then provide a detailed discussion of the relation between stochastic and mean-field results for the diffusion-only (see Sec.~\ref{secDif}) and reaction-only (see Sec.~\ref{secReac}) systems. We derive analytic solutions of the ME for special cases of the reaction dynamics at synaptic domains, and carry out extensive kinetic Monte Carlo (KMC) simulations of the ME for
the diffusion-only and reaction-only systems. Allowing for an interplay between reaction and diffusion processes at the cell membrane we explore, in Sec.~\ref{secRSD},
collective fluctuations in synaptic domains \cite{Ribrault2011,Choquet2013}, the molecular turnover at synaptic domains measured in fluorescence recovery after photobleaching (FRAP) experiments \cite{Choquet2003,Specht2008,Calamai2009}, and the stochastic single-molecule dynamics at synaptic domains \cite{Choquet2003,Choquet2013,Meier2001,Borgdorff2002,Dahan2003,Specht2008,Triller2005,Triller2008}. We conclude, in Sec.~\ref{secSum}, with a summary and discussion of our key results. Appendices~\ref{Appendix A} and~\ref{Appendix B} provide mathematical
details pertaining to our analytic solutions of the ME for the reaction-only~system.

\section{Reaction-diffusion model of synaptic receptor-scaffold domains}
\label{secModel}

\begin{figure}[t!]
\includegraphics[width=\columnwidth]{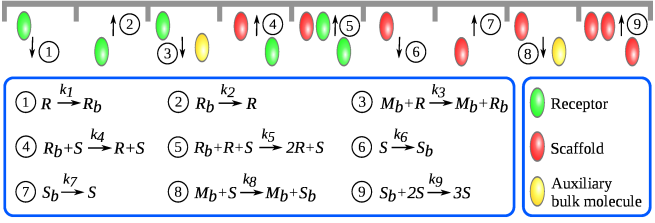} 
\caption{Schematic of the stochastic lattice model of the reaction dynamics of synaptic receptor and scaffold molecules \cite{Haselwandter2011,Haselwandter2015,Kahraman2016}
considered in this article. To model molecular crowding, the rates of all reaction and diffusion processes increasing the molecule number at a given lattice site $i$, delineated by vertical ticks, are taken to be $\propto \left(1-N^r_i-N^s_i \right)$. The transition rates associated with the indicated reaction processes are given by Eq.~(\ref{eq:transition}) with Eqs.~(\ref{eq:ReacTerm1})--(\ref{eq:ReacTerm9}).
}
\label{fig:schematic}
\end{figure}

In this section we summarize the stochastic lattice model of the reaction and diffusion dynamics of synaptic receptor ($R$) and scaffold ($S$) molecules developed in Refs.~\cite{Haselwandter2011,Haselwandter2015,Kahraman2016},
which we use throughout this article (see Fig.~\ref{fig:schematic}). In this model, the membrane is discretized into membrane patches (lattice sites). We assume that chemical reactions only take place among receptors or scaffolds occupying the same lattice site, with random hopping of receptors and scaffolds between nearest-neighbor lattice sites. We focus here on the most straightforward scenario of a 1D system of length $L$ with $K$ patches of size $a=L/K$. The 2D formulation of our model \cite{Haselwandter2011,Haselwandter2015} shows
\cite{Kahraman2016} similar stochastic dynamics of synaptic domains as the 1D formulation we consider here. We denote the hopping rates of receptors and scaffolds at lattice site $i$ by $D_i^{r}/\tau_r$ and $D_i^{s}/\tau_s$, where $D_i^{r}(t)$ and $D_i^{s}(t)$ model spatiotemporal variations in the receptor and scaffold hopping rates.

Synaptic membrane domains are crowded with molecules \cite{Choquet2013,Ziv2014}, which is expected \cite{Triller2005,Specht2008,Triller2008,Ribrault2011,Salvatico2015} to affect diffusion and reaction processes at synaptic domains. To account for molecular crowding in our model, we impose \cite{Haselwandter2011,Haselwandter2015,Kahraman2016}
the constraint that the rates of all reaction and diffusion processes that increase the receptor or scaffold number at a lattice site $i$ are $\propto \left(1-N^r_i-N^s_i \right)$, where $N^r_i/\epsilon^r$ and $N^s_i/\epsilon^s$ are the occupation numbers of receptors and scaffolds at site $i$ with the normalization constants $\epsilon^r$ and $\epsilon^s$ so that, at each site, the number of receptors and scaffolds cannot increase beyond $1/\epsilon^r$ and $1/\epsilon^s$, respectively. As a result, we have $0\leqslant N^r_i+N^s_i\leqslant 1$ for all $i$. Analogous phenomenological models of crowding have
been employed previously in a variety of different contexts \cite{Satulovsky1996,McKane2004,Lugo2008,Fanelli2010,Fanelli2013}.
Based on recently-developed computational methodologies 
\cite{Flegg2012,Schoneberg2013,Gillespie2014,Johnson2014,Vijaykumar2015,Metzler2016,Jeon2016}
for the description of reaction and diffusion processes at molecular scales,
the simple model of crowding we focus on here could be connected to more detailed molecular models of the interactions between receptors and scaffolds.

In our stochastic reaction-diffusion model of synaptic domains \cite{Haselwandter2011,Haselwandter2015,Kahraman2016},
the state of the system at time $t$ is completely characterized by the set of molecular occupation numbers $\mathbf{N}=\{\mathbf{N}^{\alpha}\}$ with $\alpha=r,s$, where $\mathbf{N}^{\alpha}(t)=(N^\alpha_1(t),N^\alpha_2(t),\cdots,N^\alpha_K(t))$. The stochastic dynamics of the system are governed by the ME~\cite{Gardiner1985,Kampen1992}
\begin{equation}
\frac{\partial P}{\partial t}=\sum_{{\bf m}} \!\,\bigl[W({\bf N}-{\bf m};{\bf m}) P({\bf N}-{\bf m},t) - W({\bf N};{\bf m})P({\bf N},t)\bigr]  \,,
\label{eq:ME}
\end{equation}
where $P(\mathbf{N},t)$ is the probability that the system is in state $\mathbf{N}$ at time $t$ and $W(\mathbf{N};\mathbf{m})$ is the transition rate from state $\mathbf{N}$ to state $\mathbf{N}+\mathbf{m}$. Unless indicated otherwise,
we use here \cite{Haselwandter2011,Haselwandter2015,Kahraman2016} random initial conditions of $\mathbf{N}$ satisfying $0 \leq N^r_i + N^s_i \leq 1$ for all $i$ with periodic boundary conditions.

The total transition rate $W$ in Eq.~(\ref{eq:ME}) can be written~as
\begin{equation}
W=W_\text{react}+W_\text{diff}
\end{equation}
in our reaction-diffusion
model of synaptic domains \cite{Haselwandter2011,Haselwandter2015,Kahraman2016},
where $W_\text{react}$ and $W_\text{diff}$ denote contributions to $W$ due to receptor and scaffold reaction and diffusion processes at the cell membrane.
For the receptor and scaffold diffusion processes we have
\begin{equation}
W_\text{diff}=W_\text{diff}^r+W_\text{diff}^s
\end{equation}
with $W_\text{diff}^\alpha
=W_\text{diff}^{(1;\alpha)}+W_\text{diff}^{(2;\alpha)}$, in which the $W_\text{diff}^{(1,2;\alpha)}$ denote the receptor and scaffold transition rates for hopping from site $i$ to site $i\pm1$,
\begin{align} 
W_\text{diff}^{(1,2;\alpha)}(\mathbf{N};\mathbf{m}) &= \frac{1}{2 \epsilon^{\alpha} \tau_\alpha} \sum_i D_i^{\alpha}N^\alpha_i (1- N^r_{i\pm1}- N^s_{i\pm1})\nonumber
\\ &\delta\left(m_i + \epsilon^{\alpha}\right)
\delta\left(m_{i\pm1} - \epsilon^{\alpha}\right) \prod_{j\neq i,i\pm1} \delta\left(m_j\right)\,,
\nonumber\\ \label{eq:diffEx}
\end{align}
where the summation runs over the entire system, $\delta(x)$ is the Dirac-delta function, and the term $(1- N^r_{i\pm1}- N^s_{i\pm1})$ captures the effects
of molecular crowding on receptor and scaffold diffusion \cite{Haselwandter2011,Haselwandter2015}. We use Dirac-delta functions, rather than Kronecker-delta functions, in Eq.~(\ref{eq:diffEx})
in order to make the connection between the ME~(\ref{eq:ME}) and the corresponding mean-field equations more transparent, which amounts to replacing the summation in the ME~(\ref{eq:ME}) by an integral over all (continuous) ${\bf m}$ \cite{Haselwandter2007}.
The factor of $1/\epsilon^\alpha$ in Eq.~(\ref{eq:diffEx}) arises because we follow here the convention \cite{Kahraman2016} that $D_i^{\alpha}/\tau_\alpha$ is the hopping rate per molecule.

The contribution to $W$ due to reactions is given by
\begin{equation}
W_\text{react}=\sum_l W_\text{react}^{(l)}\,,
\end{equation}
in which each $W_\text{react}^{(l)}$ corresponds to a particular reaction among receptors or scaffolds. The $W_\text{react}^{(l)}$ take the general form
\begin{align} \label{eq:transition}
 W_\text{react}^{(l)}(\mathbf{N};\mathbf{m}) = \sum_i \mathcal{R}_i^{(l)} \prod_{j\neq i} \delta\left(m_j\right)\,,
\end{align}
where the summation runs over the entire system and, as in Eq.~(\ref{eq:diffEx}), we use Dirac-delta functions so as to allow for continuous ${\bf m}$ in the ME~(\ref{eq:ME}). The $\mathcal{R}_i^{(l)}$ are dictated by the receptor or scaffold reaction dynamics \cite{Specht2008,Choquet2003,Triller2008,Triller2005,Haselwandter2011,Haselwandter2015},
and we return to their specific forms below.

To derive the mean-field equations associated with our stochastic lattice
model \cite{Haselwandter2015} we introduce the continuum representations $R_i(t)$ and $S_i(t)$ of $N^r_i(t)$ and $N^s_i(t)$, respectively. Based on Eqs.~(\ref{eq:diffEx}) and~(\ref{eq:transition}), the transition rates in our reaction-diffusion model can be directly extended to the continuous occupation numbers $R_i(t)$ and $S_i(t)$, allowing the ME~(\ref{eq:ME}) to be transformed \cite{Kampen1992,Haselwandter2007,Fox1991,Horsthemke1977} into the more tractable lattice Langevin equations~\cite{Haselwandter2015}
\begin{eqnarray}
 \frac{dR_i}{dt} &=& K_i^{(r;1)} + \eta_i^{(r)} \,, \label{eq:LangevinR} \\
 \frac{dS_i}{dt} &=& K_i^{(s;1)} + \eta_i^{(s)} \,, \label{eq:LangevinS}
\end{eqnarray}
where the $K_i^{(\alpha;1)}$ are the first moments of the contributions to $W$ changing the receptor or scaffold distribution in the system, and the Gaussian noise terms $\eta_i^{(\alpha)}$ have zero mean and covariance
\begin{equation}  \label{discov}
\langle \eta_{i}^{(\alpha)}(t_1) \; \eta_{j}^{(\alpha)}(t_2)\rangle=
K_{i,j}^{(\alpha;2)}\,\delta(t_1-t_2)\, ,
\end{equation}
in which the $K_{i,j}^{(\alpha;2)}$ are the second moments of the contributions
to $W$ changing the receptor or scaffold distribution in the system.

The continuum limit of the deterministic parts of the lattice Langevin equations~(\ref{eq:LangevinR})
and~(\ref{eq:LangevinS}) yields \cite{Haselwandter2011,Haselwandter2015} the mean-field equations
\begin{eqnarray} \label{eq:MFE_r}
\frac{\partial r}{\partial t}=F^r(r,s)-\nu_r\mathbf{\nabla}\cdotp\mathbf{J}^r\,, \\
\frac{\partial s}{\partial t}=F^s(r,s)-\nu_s\mathbf{\nabla}\cdotp\mathbf{J}^s\,, \label{eq:MFE_s}
\end{eqnarray}
with all parameters determined directly by the ME~(\ref{eq:ME}), where $r(x,t)$ and $s(x,t)$ are the continuum fields associated with $R_i(t)$ and $S_i(t)$
in  Eqs.~(\ref{eq:LangevinR}) and~(\ref{eq:LangevinS}) with the noise terms
set to zero, 
\begin{align} \label{eq:continuum_limit}
Q_{i\pm n}(t) = \sum\limits_{k=0}^{\infty} \frac{\partial^{k} q}{\partial x^k}\bigg|_{x=i a}\frac{(\pm an)^k}{k!} \,,
\end{align}
in which $Q_i\equiv R_i,S_i$ and $q\equiv r(x,t),s(x,t)$, the polynomials $F^{\alpha}(r,s)$ in Eqs.~(\ref{eq:MFE_r}) and~(\ref{eq:MFE_s}) capture chemical reactions among receptors or scaffolds as in the standard formalism of chemical dynamics 
\cite{Butler2011,Cao2014,Lugo2008,Erban2009,Butler2009,Hecht2010,Erban2007,Cohen2005,Wylie2006,Samoilov2006,McKane2004,McKane2005,Gillespie1976,Gillespie1977,Gillespie2013,Epstein1998,Walgraef1997,Cross2009,Cross1993,Murray2002,Meinhardt1982,Maini2001}, the $\nu_{\alpha}= a^2/2 \tau_{\alpha}$ are the receptor and scaffold diffusion coefficients, and the diffusion currents are given by
\begin{eqnarray}
\mathbf{J}^r &=& -D^r \left(1-s\right)\nabla{r}-D^r
r\nabla{s} -\left(1-r -s\right)r\nabla{D^r}\,, \nonumber \\ \label{eq:dif_curR}
&& \\
\mathbf{J}^s &=& -D^s \left(1-r\right)\nabla{s}-D^s s\nabla{r}
-\left(1-r-s\right)s\nabla{D^s}\,,\nonumber \\ &&\label{eq:dif_curS}
\end{eqnarray}
where the $D^{\alpha}(x,t)$ denote the continuum representations \cite{Haselwandter2015}
of $D^{\alpha}_i$ obtained via Eq.~(\ref{eq:continuum_limit}) with $Q_i\equiv
D_i^{\alpha}$ and $q \equiv D^{\alpha}(x,t)$. The diffusion currents in Eqs.~(\ref{eq:dif_curR}) and~(\ref{eq:dif_curS}) follow directly \cite{Haselwandter2015} from the random hopping of receptors and scaffolds with rates $D^{\alpha}_i/\tau_{\alpha}$ together with the constraint that the rates of diffusion processes locally increasing the molecule occupancy at a given site $i$ are $\propto \left(1-N^r_i-N^s_i\right)$,
as captured by Eq.~(\ref{eq:diffEx}). Nonlinear crowding terms equivalent to those in Eqs.~(\ref{eq:dif_curR}) and~(\ref{eq:dif_curS}) have been studied previously in population biology \cite{Satulovsky1996,McKane2004,Lugo2008} and in the context of general models of non-Fickian diffusion~\cite{Fanelli2010,Fanelli2013}.
Membrane-bound reaction-diffusion systems similar to Eqs.~(\ref{eq:MFE_r}) and~(\ref{eq:MFE_s}) occur in a variety of different contexts \cite{Haselwandter2011,Haselwandter2015,Hecht2010,Levine2005,Marenduzzo2013,Vandin2016,Camley2017}.

We use glycine receptors and gephyrin scaffolds \cite{Ziv2014,Salvatico2015,Ribrault2011,Choquet2013,Specht2008,Triller2008,Triller2005}
as a model system to fix the reaction kinetics and diffusion coefficients in our reaction-diffusion model of synaptic receptor-scaffold domains. We summarize here the pertinent reaction-diffusion dynamics, and refer the interested reader to Refs.~\cite{Haselwandter2011,Haselwandter2015,Kahraman2016} for a more detailed discussion of how the reaction-diffusion model considered here relates to the experimental phenomenology of glycine receptors and gephyrin.
We first note that, at the lowest order, receptors and scaffolds may be randomly removed from the cell membrane via endocytosis as well as randomly inserted into the cell membrane, resulting in the reactions $R\xrightarrow{k_1} R_b$, $R_b\xrightarrow{k_2} R$, $S\xrightarrow{k_6} S_b$, and $S_b\xrightarrow{k_7} S$ (Fig.~\ref{fig:schematic}). In these expressions, $R$ and $S$ represent receptor and scaffold molecules at the membrane, while  $R_b$ and $S_b$ stand for receptor and scaffold molecules in the cytoplasmic ``bulk'' of the cell \cite{Haselwandter2011,Haselwandter2015}, with the $k_l$ denoting rate constants. The resulting transition rates are given by Eq.~(\ref{eq:transition}) with \cite{Haselwandter2015,Kahraman2016}
\begin{eqnarray}
\mathcal{R}_i^{(1)} &=& \frac{k_1}{\epsilon^r} N^r_i  \delta\left(m_i + \epsilon^r\right) \, , \label{eq:ReacTerm1}
 \\ \mathcal{R}_i^{(2)} &=& \frac{k_2}{\epsilon^r} (1-N^r_i-N^s_i) \delta\left(m_i - \epsilon^r\right)\, , \label{eq:ReacTerm2} \\
 \mathcal{R}_i^{(6)} &=&  \frac{k_6}{\epsilon^s} N^s_i  \delta\left(m_i + \epsilon^s\right)\, , \label{eq:ReacTerm3} \\
 \mathcal{R}_i^{(7)} &=& \frac{k_7}{\epsilon^s} (1-N^r_i-N^s_i) \delta\left(m_i - \epsilon^s\right) \, , \label{eq:ReacTerm4} 
\end{eqnarray}
yielding the additive contributions $-k_1 r$ and $k_2 (1-r-s)$ to $F^r$ in Eq.~(\ref{eq:MFE_r}), and $-k_6 s$ and $k_7 (1-r-s)$ to $F^s$ in Eq.~(\ref{eq:MFE_s}).
The rate constants $k_l$ in Eqs.~(\ref{eq:ReacTerm1})--(\ref{eq:ReacTerm4})
[as well as the rate constants in Eqs.~(\ref{eq:ReacTerm5})--(\ref{eq:ReacTerm9});
see below] are scaled by $1/\epsilon^\alpha$ because we use the convention
\cite{Kahraman2016} that $k_l$ denotes the rate of removal from/insertion into the cell membrane per molecule.

Furthermore, we note \cite{Haselwandter2011,Haselwandter2015,Kahraman2016}
that removal of receptors or scaffolds from the cell membrane may be facilitated by some mechanism that involves a temporary increase in the local crowding of the cell membrane, $M_b+R\xrightarrow{k_3} M_b+R_b$ and $M_b+S\xrightarrow{k_8} M_b+S_b$ (Fig.~\ref{fig:schematic}), where $M_b$ denotes an auxiliary bulk molecule. Because we account here for the effects of molecular crowding, these reactions yield contributions to the stochastic lattice model that are distinct from $R\xrightarrow{k_1} R_b$ and $S\xrightarrow{k_6} S_b$, and result in the terms \cite{Haselwandter2015,Kahraman2016}
\begin{eqnarray}
 \mathcal{R}_i^{(3)} &=& \frac{k_3}{\epsilon^r} (1-N^r_i-N^s_i) N^r_i \delta\left(m_i + \epsilon^r\right)\, , \label{eq:ReacTerm5} \\
 \mathcal{R}_i^{(8)} &=& \frac{k_8}{\epsilon^s} (1-N^r_i-N^s_i) N^s_i \delta\left(m_i + \epsilon^s\right) \label{eq:ReacTerm6} 
\end{eqnarray}
in Eq.~(\ref{eq:transition}). Equations~(\ref{eq:ReacTerm5}) and~(\ref{eq:ReacTerm6})
imply the additive contributions  $-k_3(1-r-s)r$ to $F^r$ in Eq.~(\ref{eq:MFE_r}) and $-k_8(1-r-s)s$ to $F^s$ in Eq.~(\ref{eq:MFE_s}), respectively.

Finally, we note that, as discussed previously \cite{Haselwandter2011,Haselwandter2015}, key experimental features of the reaction dynamics of glycine receptors and gephyrin for self-assembly of synaptic domains \cite{Kirsch1995,Meier2000,Meier2001,Borgdorff2002,Dahan2003,Hanus2006,Ehrensperger2007,Calamai2009,Haselwandter2011}
are \cite{Specht2008,Choquet2003,Triller2008,Triller2005,Haselwandter2011,Haselwandter2015}
that gephyrin can transiently bind glycine receptors as well as other gephyrin molecules, with experiments and theory suggesting \cite{Choquet2003,Calamai2009,Kahraman2016,Haselwandter2011,Haselwandter2015}
that trimerization of gephyrin is a key reaction for self-assembly of synaptic domains. Allowing for the same order of receptor reactions as scaffold
reactions, these considerations suggest \cite{Haselwandter2015} the reactions $R_b+S\xrightarrow{k_4} R+S$, $R_b+R+S\xrightarrow{k_5} 2R+S$, and $S_b+2S\xrightarrow{k_9} 3S$ (Fig.~\ref{fig:schematic}), resulting in the terms \cite{Haselwandter2015,Kahraman2016}
\begin{align}
 \mathcal{R}_i^{(4)} =&~ \frac{k_4}{\epsilon^r} (1-N^r-N^s) N^s \delta\left(m_i - \epsilon^{r}\right)\, , \label{eq:ReacTerm7} \\
 \mathcal{R}_i^{(5)} =&~ \frac{k_5}{\epsilon^r} (1-N^r-N^s) N^r N^s \delta\left(m_i - \epsilon^{r}\right)\, , \label{eq:ReacTerm8} \\
 \mathcal{R}_i^{(9)} =&~ \frac{k_9}{2!\epsilon^s} (1-N^r-N^s) N^s (N^s-\epsilon^s) \delta\left(m_i - \epsilon^{s}\right) \label{eq:ReacTerm9}
\end{align}
in Eq.~(\ref{eq:transition}). Equations~(\ref{eq:ReacTerm7}) and~(\ref{eq:ReacTerm8}) yield the additive contributions $k_4(1-r-s)s$ and $k_5(1-r-s)r s$ to
$F^r$ in Eq.~(\ref{eq:MFE_r}), and Eq.~(\ref{eq:ReacTerm9}) implies the additive
contribution $k_9(1-r-s)s^2/2$ to $F^s$ in Eq.~(\ref{eq:MFE_s}).

\begin{table}[t!]
\caption{Unless indicated otherwise, we use here the same reaction kinetics and values of the dimensionless rate constants as in Refs.~\cite{Haselwandter2011,Kahraman2016}, which correspond to \textit{model C} in Ref.~\cite{Haselwandter2015} and are consistent with experiments on glycine receptors and gephyrin scaffolds \cite{Specht2008,Choquet2003,Triller2008,Triller2005}: $(m_1,m_2,\beta,\mu)=b(0.4,10,0.5,0.7)$ and $(\bar{r},\bar{s})=(0.05,0.05)$, with the right column in the table showing the connection between the notation used here and in Refs.~\cite{Haselwandter2011,Haselwandter2015}.
As in Ref.~\cite{Kahraman2016}, we fix the time units in our model by adjusting the rate of receptor endocytosis within the range of values estimated previously \cite{Haselwandter2011,Haselwandter2015} from experiments, which correspond
to characteristic time scales ranging from seconds to hours \cite{Specht2008,Choquet2003,Triller2008}, to $k_1=1/750$~$\text{s}^{-1}$ so that our model reproduces the scaffold recovery time measured in FRAP experiments \cite{Choquet2003,Specht2008,Calamai2009}.
The indicated rate constants enter our (stochastic and mean-field) reaction-diffusion model of synaptic domains through Eq.~(\ref{eq:transition}) with Eqs.~(\ref{eq:ReacTerm1})--(\ref{eq:ReacTerm9}).}
\label{tabReactionRates}
\begin{ruledtabular}
\begin{tabular}{ccc} \vspace{0.05cm}
Chemical reactions & Rate constants \\ \hline \\ [-0.4cm]
$R\xrightarrow{k_1} R_b$      & $k_1=b \approx$~$1.3\times 10^{-3}$~s$^{-1}$\\
$R_b\xrightarrow{k_2} R$         & $k_2=m_1 \frac{\bar{r}}{1-\bar{r}-\bar{s}}\approx
3.0\times10^{-5}$~s$^{-1}$\\
$M_b+R\xrightarrow{k_3} M_b+R_b$ & $k_3=\frac{m_1\bar{r}+m_2\bar{s}}{\bar{r}(1-\bar{r}-\bar{s})}\approx
1.5\times10^{-2}$~s$^{-1}$\\
$R_b+S\xrightarrow{k_4} R+S$     & $k_4=b\frac{\bar{r}}{\bar{s}}\frac{1}{1-\bar{r}-\bar{s}}\approx
1.5\times 10^{-3}$~s$^{-1}$\\
$R_b+R+S\xrightarrow{k_5} 2R+S$  & $k_5=\frac{m_2}{\bar{r}}\frac{1}{1-\bar{r}-\bar{s}}\approx
3.0 \times10^{-1}$~s$^{-1}$\\ $S\xrightarrow{k_6} S_b$ & $k_6=\beta\approx 6.7\times10^{-4}$~s$^{-1}$ \\
$S_b\xrightarrow{k_7} S$ & $k_7=\beta \frac{\bar{s}}{1-\bar{r}-\bar{s}}\approx
3.7\times 10^{-5}$~s$^{-1}$\\
$M_b+S\xrightarrow{k_8} M_b+S_b$ & $k_8=\frac{\mu}{1-\bar{r}-\bar{s}}\approx
1.0\times10^{-3}$~s$^{-1}$\\
$S_b+2S\xrightarrow{k_9} 3S$ & $k_9=\frac{\mu}{\bar{s}}\frac{2}{1-\bar{r}-\bar{s}}\approx
4.1\times 10^{-2}$~s$^{-1}$
\end{tabular}
\end{ruledtabular}
\end{table}

Unless indicated otherwise, we use here the same values of $k_l$ as in Ref.~\cite{Kahraman2016} (see Table~\ref{tabReactionRates}) which, as discussed in Refs.~\cite{Haselwandter2011,Haselwandter2015,Kahraman2016}, are consistent with experiments on synaptic domains formed by glycine receptors
and gephyrin scaffolds. Similarly, we use \cite{Kahraman2016}, unless indicated otherwise, the diffusion coefficients $\nu_r=10^2\nu_s=10^{-2} \mu \text{m}^2/\text{s}$,
with the corresponding hopping rates $1/\tau_{\alpha}=2 \nu_{\alpha}/a^2$ in Eq.~(\ref{eq:ME}), consistent with experiments on glycine receptors and gephyrin scaffolds \cite{Specht2008,Choquet2003,Triller2008,Triller2005,Calamai2009,Meier2001,Haselwandter2011,Haselwandter2015}.
For simplicity, we set $\epsilon^r = \epsilon^s \equiv \epsilon$ and $D^r=D^s=1$ throughout this article. Distinct values of $\epsilon^r$ and $\epsilon^s$ could be used to provide a more detailed model of the receptor and scaffold numbers at synaptic domains, while spatiotemporal variations in $D^r$ or $D^s$ could be used \cite{Haselwandter2011,Haselwandter2015}, for instance, to model the effects of pre- and postsynaptic interactions on receptor or scaffold diffusion. Unless indicated otherwise, we set $\epsilon=1/100$ and $a\approx 80$~nm so that \cite{Erban2009,Cao2014} the membrane patch size is smaller than the expected typical size of synaptic domains \cite{Kirsch1995,Meier2000,Meier2001,Borgdorff2002,Dahan2003,Hanus2006,Ehrensperger2007,Calamai2009,Haselwandter2011,Haselwandter2015}
but large enough to accommodate multiple receptors and scaffolds, with size $\approx5$--$10$~nm for glycine receptors and gephyrin \cite{Kim2006,Du2015}.

As discussed in Sec.~\ref{secReac}, we obtained exact analytic solutions
of the ME~(\ref{eq:ME}) for reaction-only systems involving a subset of the reactions in Eqs.~(\ref{eq:ReacTerm1})--(\ref{eq:ReacTerm9}). We supplemented these exact analytic solutions for general reaction schemes through direct
numerical solutions of the ME~(\ref{eq:ME}), for which we used the Euler method. Furthermore, we carried out KMC simulations of the ME~(\ref{eq:ME}) \cite{Gillespie1976,Gillespie1977,Gillespie2013} for diffusion-only (see Sec.~\ref{secDif}), reaction-only (see Sec.~\ref{secReac}), and reaction-diffusion (see Sec.~\ref{secRSD}) systems employing the ``spatial next reaction'' method described in Ref.~\cite{Elf2003}. In our implementation of the spatial next reaction method \cite{Elf2003} we used Gillespie's ``direct'' method \cite{Gillespie1976} to choose, at each lattice site, which receptors or scaffolds undergo reaction or hopping processes. On this basis we were able to track individual receptors and scaffolds in our KMC simulations. Finally, we numerically solved the mean-field equations~(\ref{eq:MFE_r}) and~(\ref{eq:MFE_s}) using standard methods \cite{Mathematica} with the initial conditions, boundary conditions, and parameter values employed for the ME~(\ref{eq:ME}).

\section{Protein diffusion in crowded membranes}
\label{secDif}

In this section we focus on diffusion-only systems described by the ME~(\ref{eq:ME}) with $W_\textrm{react}=0$ and $D^r_i=D^s_i=1$. For such systems, Eqs.~(\ref{eq:MFE_r})
and~(\ref{eq:MFE_s}) imply the mean-field diffusion equations \cite{Satulovsky1996,McKane2004,Lugo2008,Fanelli2010,Fanelli2013,Haselwandter2011,Haselwandter2015}
\begin{eqnarray} \label{eq:MFE_difonlyR}
 \frac{\partial r}{\partial t}&=&\nu_r\left[\left(1-s\right)\nabla^2r+r\nabla^2s\right]\,, \\ 
 \frac{\partial s}{\partial t}&=&\nu_s\left[\left(1-r\right)\nabla^2s+s\nabla^2r\right]\,. \label{eq:MFE_difonlyS}
\end{eqnarray}
The nonlinear terms in the mean-field equations~(\ref{eq:MFE_difonlyR}) and~(\ref{eq:MFE_difonlyS}) result from molecular crowding (steric exclusion), and impede diffusion
into crowded membrane regions. In line with experiments and large-scale computer simulations of crowded membranes \cite{Metzler2016,Jeon2016}, the nonlinear diffusion terms in Eqs.~(\ref{eq:MFE_difonlyR}) and~(\ref{eq:MFE_difonlyS}) have been shown \cite{Fanelli2010} to result in mean-square displacement curves that bear signatures of anomalous diffusion. Below, we first consider the special case $\nu_r=\nu_s\equiv \nu$ in Eqs.~(\ref{eq:MFE_difonlyR}) and~(\ref{eq:MFE_difonlyS}), for which the total molecule concentration of
receptors and scaffolds, $r+s$, obeys the standard (linear) diffusion equation
\begin{equation} \label{eq:DiffTotal}
\frac{\partial (r+s)}{\partial t}= \nu \nabla^2 (r+s)\,,
\end{equation}
and then discuss more complex scenarios corresponding to $\nu_r \neq \nu_s$.

\subsection{Identical receptor and scaffold diffusion coefficients}
\label{secIdentical}

\begin{figure}[t!]
\includegraphics[width=\columnwidth]{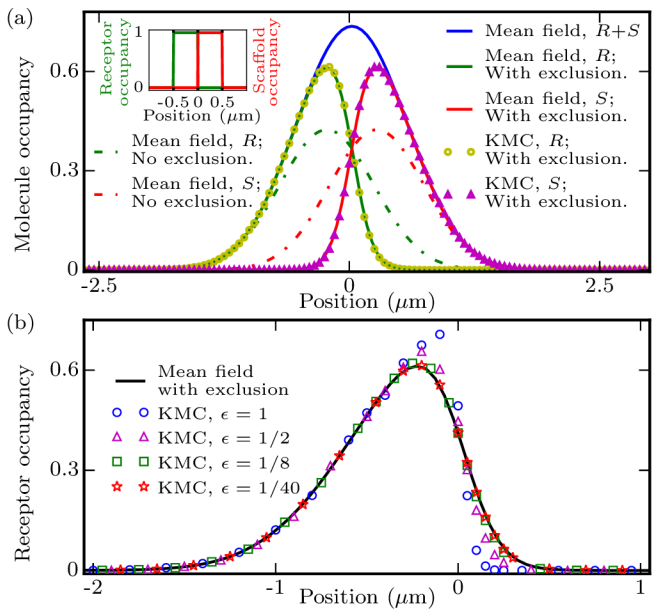} 
\caption{Receptor and scaffold diffusion with $\nu_r=\nu_s\equiv\nu=0.01$~$\mu$m$^2/$s. (a) Molecule occupancies of receptors and scaffolds at $t=10$~s starting from the initial conditions shown in the inset, obtained from KMC simulations of the ME~(\ref{eq:ME}) with $\epsilon=1/40$ and $a=0.05$~$\mu$m, the mean-field equations~(\ref{eq:MFE_difonlyR}) and (\ref{eq:MFE_difonlyS}) modeling diffusion under exclusion constraints, and the standard (Fickian) diffusion equations for receptors and scaffolds, which are given by the linear terms in Eqs.~(\ref{eq:MFE_difonlyR}) and (\ref{eq:MFE_difonlyS}). We set $L=50$~$\mu$m. (b) Receptor profiles as in panel (a), but using $\epsilon=1$, 1/2, 1/8, and 1/40 for the KMC simulations of the ME~(\ref{eq:ME}). The KMC simulations were averaged over 2000 independent realizations each.}
\label{fig:dif1}
\end{figure}

\begin{figure}[t!]
\includegraphics[width=\columnwidth]{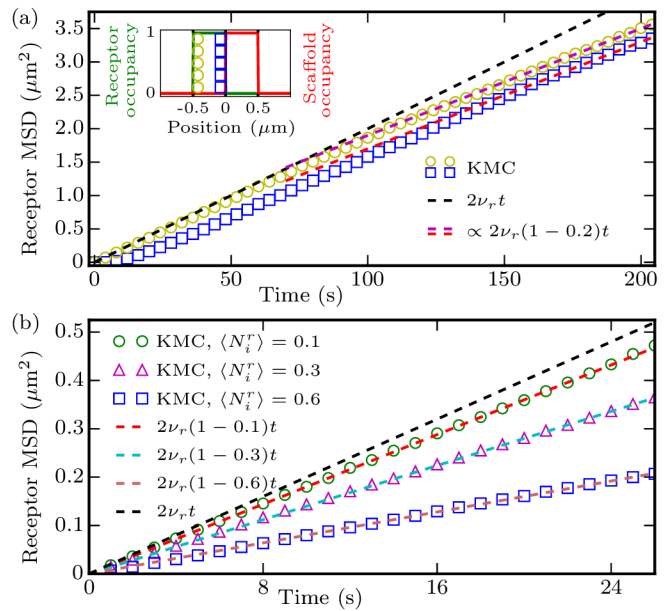} 
\caption{MSD in diffusion-only systems with $\nu_r=\nu_s\equiv\nu=0.01$~$\mu$m$^2/$s. (a) Receptor MSD obtained from KMC simulations as in Fig.~\ref{fig:dif1}(a)
but using $L=5$~$\mu$m, for receptors located initially at the positions marked by circles and squares in the inset. The black dashed line shows the MSD implied by standard (Fickian) diffusion with no steric constraints, and the red and magenta dashed lines show the MSD for free diffusion scaled by $\left(1-\langle N_i^r + N_i^s \rangle\right)$, where $\langle N_i^r + N_i^s \rangle=0.2$ is the average molecule occupancy in the system. (b) Receptor MSD obtained
as in panel (b), but for a system composed of only receptors and starting from homogeneous receptor distributions with $\langle N_i^r \rangle=0.1$, $0.3$, and $0.6$. The MSD curves were obtained by averaging over $10^4$ molecule trajectories each.
}
\label{fig:dif1b}
\end{figure}

\begin{figure*}[t!]
\center
\includegraphics[width=\textwidth]{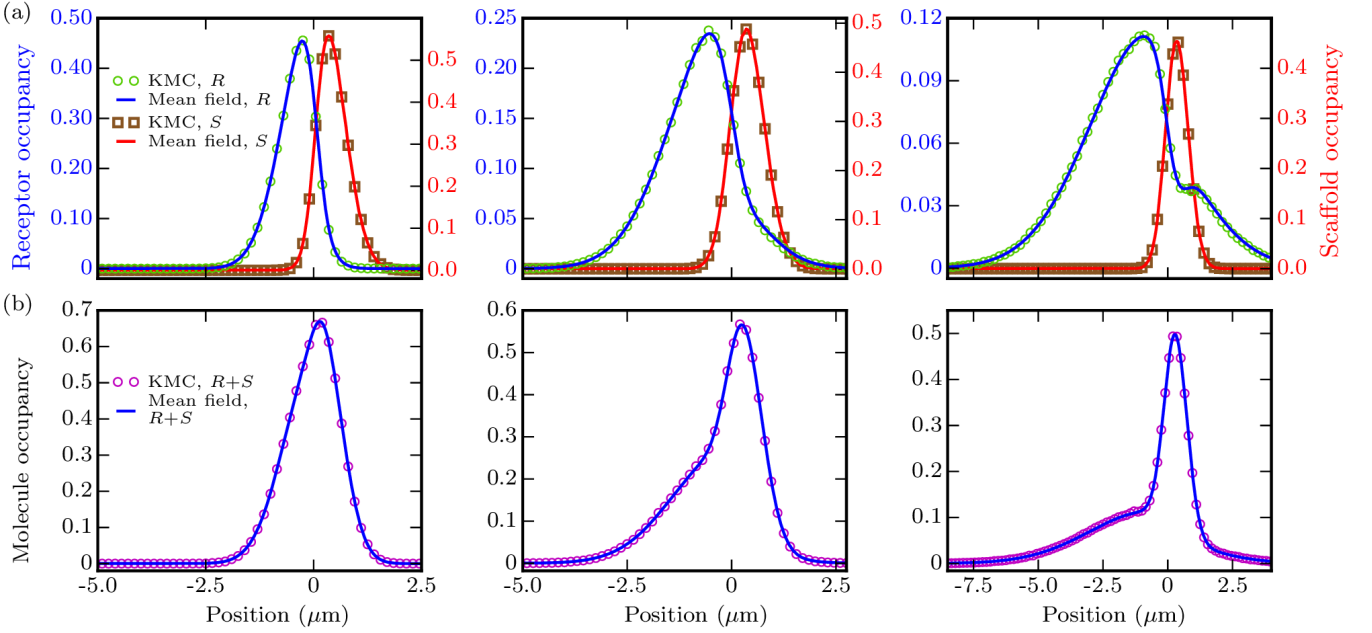} 
\caption{Profiles of (a) receptor and scaffold occupancies and (b) the
total receptor and scaffold occupancy, obtained by adding up the curves
in panel (a), at $t=10$~s for $\nu_r/\nu_s=2,8,32$ for the left, middle, and right panels, respectively, calculated from KMC simulations of the ME~(\ref{eq:ME}) and the mean-field equations~(\ref{eq:MFE_difonlyR}) and (\ref{eq:MFE_difonlyS}) as in Fig.~\ref{fig:dif1}(a) using $\nu_s=0.01$~$\mu$m$^2$/s and $\epsilon=1/40$. The KMC simulations were averaged over 2000 independent realizations each.}
\label{fig:dif2}
\end{figure*}

\begin{figure}[t!]
\includegraphics[width=\columnwidth]{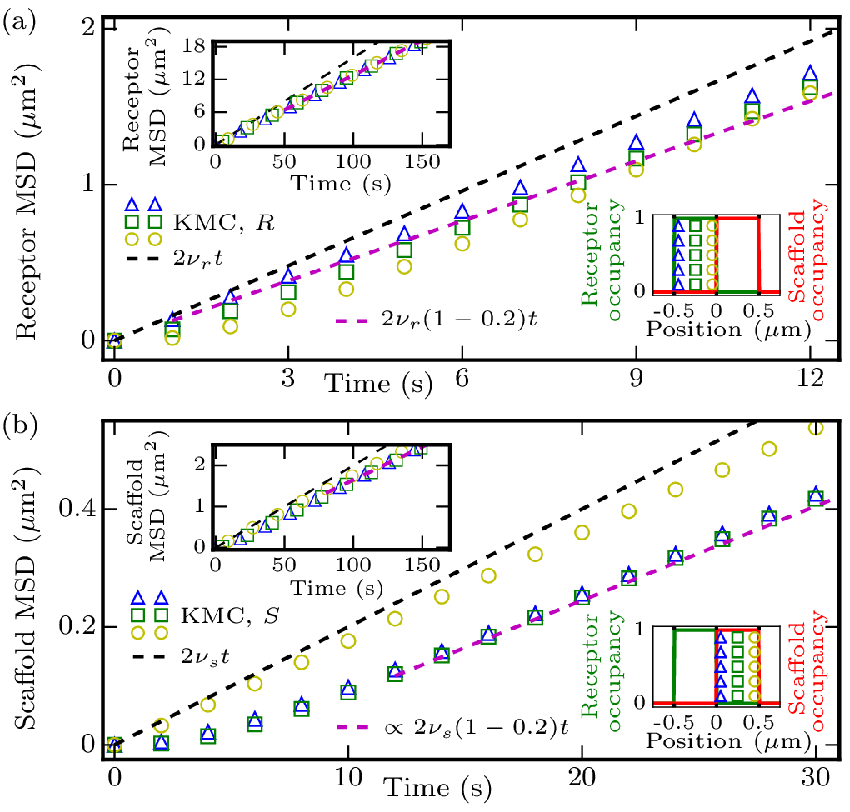} 
\caption{MSD curves of (a) diffusing receptors and (b) diffusing scaffolds obtained from KMC simulations as in Fig.~\ref{fig:dif1}(a) with $\epsilon=1/40$
but using $L=5$~$\mu$m
with $\nu_r=0.08$~$\mu$m$^2$/s and $\nu_s=0.01$~$\mu$m$^2$/s, for receptors and scaffolds located initially at the positions marked by triangles, squares,
and circles in the initial molecule distributions shown in the lower-right insets. The upper-left insets show the long-term evolution of the MSD curves. As in Fig.~\ref{fig:dif1b}, the black dashed lines indicate the MSDs implied by standard (Fickian) diffusion with no steric constraints, and the magenta dashed lines show the MSDs for free diffusion scaled by $\left(1-\langle N_i^r + N_i^s \rangle\right)$, where $\langle N_i^r + N_i^s \rangle=0.2$ is the average molecule occupancy in the system. The MSD curves were obtained by averaging over $10^4$ molecule trajectories each.}
\label{fig:msd2}
\end{figure}

In this section we focus on diffusion-only systems with the diffusion coefficients of receptors and scaffolds being equal to each other, $\nu_r=\nu_s\equiv\nu=0.01$~$\mu$m$^2/$s (see Fig.~\ref{fig:dif1}). As described in Sec.~\ref{secModel}, we expect
that $\nu_r > \nu_s$ for synaptic receptors and scaffolds \cite{Triller2008,Choquet2013,Choquet2003,Triller2005,Okabe1999,Gray2006,Calamai2009,Ziv2014,Kneussel2014}, but we consider here the case $\nu_r=\nu_s$ for completeness.
As initial conditions we use adjacent step-profiles of receptors and scaffolds, with receptors and scaffolds changing from $N^{r}_i=0$
or $N^{s}_i=0$ ($r=0$ or $s=0$) to $N^{r}_i=1$ or $N^{s}_i=1$ ($r=1$ or $s=1$) [see the inset of Fig.~\ref{fig:dif1}(a)]. We find that, for small enough $\epsilon$, the mean-field equations~(\ref{eq:MFE_difonlyR}) and (\ref{eq:MFE_difonlyS}) are in quantitative agreement with averages
over KMC simulations of the underlying ME~(\ref{eq:ME}) with $W_\textrm{react}=0$ and $D^r_i=D^s_i=1$. For instance, Fig.~\ref{fig:dif1}(a)
shows excellent agreement between mean-field equations and averages over
KMC simulations for $\epsilon=1/40$. As the value $\epsilon$ is increased,
the discreteness of the molecular diffusion processes becomes increasingly important, and the mean-field approach begins to yield inaccurate results for the average molecule concentrations [see Fig.~\ref{fig:dif1}(b)]. However, we find that even with $\epsilon =1/8$, which corresponds to a maximum molecule occupancy per lattice site of only eight molecules, the mean-field equations~(\ref{eq:MFE_difonlyR}) and (\ref{eq:MFE_difonlyS}) capture the shape of the average diffusion profiles.
As expected from Eqs.~(\ref{eq:MFE_difonlyR}) and (\ref{eq:MFE_difonlyS})
with Eq.~(\ref{eq:DiffTotal}), the temporal evolution of the total molecule concentration $N^{r}_i+N^s_i$ ($r+s$) takes the form of a Gaussian profile corresponding to standard (Fickian) diffusion with diffusion coefficient $\nu$. In contrast, the individual receptor and scaffold concentration profiles do not take the form of Gaussian concentration profiles, with diffusion of receptor and scaffold populations being hindered by steric~constraints.

To explore the effect of molecular crowding on the diffusion of individual molecules, we used KMC simulations to compute the mean-square displacement (MSD) curves of receptors located at different initial positions in Fig.~\ref{fig:dif1} [see Fig.~\ref{fig:dif1b}(a)]. We find that receptors initially located at the center of a crowded membrane region show effective hopping rates that are reduced substantially compared to the case of free diffusion, resulting in a reduced initial slope of the MSD. In contrast, receptors initially located near the boundary of a crowded membrane region can easily diffuse into membrane regions with few receptors or scaffolds, resulting in an initial slope of the MSD that is only reduced slightly compared to the case of free diffusion. As $t \to \infty$, all molecules have the same effective diffusion coefficient $\nu \left(1-\langle N_i^r +N_i^s\rangle\right)$ independent of their initial location, where $\langle N_i^r +N_i^s\rangle$ is the average number of molecules per lattice site in the system. Thus, molecular crowding can initially yield spatially heterogeneous MSD curves that bear a signature of the initial location of the molecule under consideration, and asymptotically results in a reduced effective diffusion coefficient, with the magnitude of the reduction in the effective diffusion coefficient governed by the value of $\langle N_i^r +N_i^s\rangle$ [see Fig.~\ref{fig:dif1b}(b)].

\subsection{Distinct receptor and scaffold diffusion coefficients}

Synaptic receptors are thought to diffuse more rapidly than their associated scaffolds \cite{Triller2008,Choquet2013,Choquet2003,Triller2005,Okabe1999,Gray2006,Calamai2009,Ziv2014,Kneussel2014},
and we therefore focus here on the case $\nu_r > \nu_s$. Similarly as in Sec.~\ref{secIdentical} we find \cite{Kahraman2016} that, provided $\epsilon \lessapprox1/10$, the mean-field equations~(\ref{eq:MFE_difonlyR}) and~(\ref{eq:MFE_difonlyS}) describing a diffusion-only system composed of receptors and scaffolds with $\nu_r/\nu_s>1$ are in quantitative agreement with averages over KMC simulations of the underlying ME~(\ref{eq:ME}), independent of the particular value of
$\nu_r/\nu_s$ considered (see Fig.~\ref{fig:dif2}). We thus find that, even in situations where steric constraints severely limit the maximum receptor or scaffold occupancy per membrane patch and the diffusion profiles deviate
substantially from Fickian diffusion, the nonlinear mean-field diffusion
equations~(\ref{eq:MFE_difonlyR}) and~(\ref{eq:MFE_difonlyS}) \cite{Satulovsky1996,McKane2004,Lugo2008,Fanelli2010,Fanelli2013}
successfully predict the average concentration profiles implied by the underlying stochastic model, with all parameters in Eqs.~(\ref{eq:MFE_difonlyR}) and~(\ref{eq:MFE_difonlyS})
determined directly by the ME~(\ref{eq:ME}).

As $\nu_r/\nu_s$ is increased, molecular crowding produces increasingly complex receptor and scaffold concentration profiles (Fig.~\ref{fig:dif2}). In particular,
using adjacent step-profiles of receptors and scaffolds as initial conditions,
we find that the slowly-diffusing scaffolds act as a partially permeable (and dynamic) barrier to receptor diffusion that can, for large enough $\nu_r/\nu_s$, lead to non-monotonic receptor concentration profiles [see the right panel of Fig.~\ref{fig:dif2}(a)]. This can be understood by noting that, with the initial conditions used for Fig.~\ref{fig:dif2}, the slowly-varying scaffold concentration profiles show a pronounced maximum for a sustained period of time. Equation~(\ref{eq:MFE_difonlyR}) indicates that membrane regions with $\nabla^2 s<0$ locally depress the receptor concentration profile, because steric constraints make it unfavorable for receptors to diffuse into membrane regions that are crowded with scaffolds. As shown in the right panel of Fig.~\ref{fig:dif2}(a), a maximum in the scaffold concentration profile can therefore result in a local (transient) minimum in the receptor concentration profile and, hence, a multimodal receptor concentration profile. We also note that, with increasing $\nu_r/\nu_s$, the receptor profile becomes increasingly asymmetric, with an increasingly pronounced tail away from the scaffolds. As a result, the total molecule concentration profile $N^{r}_i+N^s_i$ ($r+s$) also becomes increasingly asymmetric [see Fig.~\ref{fig:dif2}(b)]. As expected from Eqs.~(\ref{eq:MFE_difonlyR}) and (\ref{eq:MFE_difonlyS}), the total molecule concentration profile $N^{r}_i+N^s_i$ ($r+s$), as well as the receptor and scaffold concentration profiles, do not take the form of Gaussian diffusion profiles if $\nu_r \neq \nu_s$.

Starting from step-like initial concentration profiles of receptors and scaffolds, we used KMC simulations to compute the MSD curves of receptors [see Fig.~\ref{fig:msd2}(a)]
and scaffolds [see Fig.~\ref{fig:msd2}(b)] located at left, center, and right positions in the initial molecule distributions of receptors and scaffolds.
Consistent with Fig.~\ref{fig:dif1b}(a), we find that receptors and scaffolds initially located at the boundaries of the crowded membrane region are least constrained by molecular crowding, and show the largest MSDs. Receptors initially located close to the center of the crowded membrane region show the smallest initial MSDs of all receptors,
because their diffusion is hindered by receptors in one direction and by
scaffolds in the other direction. In contrast, scaffolds close to the domain
center show similar initial MSDs as scaffolds initially located at the center of the scaffold distribution. This can be understood by noting that, even though the diffusion
of scaffolds initially located close to the center of the crowded membrane region is hindered by both receptors and scaffolds, the rapid diffusion of receptors soon allows scaffolds to diffuse into membrane regions that were initially fully
occupied by receptors. Similarly as in Fig.~\ref{fig:dif1b}, the asymptotic properties of the receptor and scaffold MSD curves in Fig.~\ref{fig:msd2} are set by the receptor and scaffold diffusion constants scaled by $\left(1-\langle N_i^r + N_i^s \rangle\right)$, independent of the initial location of receptors
or scaffolds. However, the crossover time from the initial to the asymptotic properties of the MSD curves is sensitive to the initial conditions used.

\section{Protein reaction dynamics in crowded membranes}
\label{secReac}

In this section we focus on reaction-only systems described by the ME~(\ref{eq:ME}) with $W_\textrm{diff}=0$, and the corresponding mean-field equations
\begin{eqnarray} \label{eq:MFE_reactonlyR}
 \frac{d r}{d t}&=&F^r(r,s)\,, \\ 
 \frac{d s}{d t}&=&F^s(r,s)\,, \label{eq:MFE_reactonlyS}
\end{eqnarray}
where, as described in Sec.~\ref{secModel}, the polynomials $F^r(r,s)$ and $F^s(r,s)$ are formulated in accordance with the standard formalism of chemical dynamics 
\cite{Butler2011,Cao2014,Lugo2008,Erban2009,Butler2009,Hecht2010,Erban2007,Cohen2005,Wylie2006,Samoilov2006,McKane2004,McKane2005,Gillespie1976,Gillespie1977,Gillespie2013,Epstein1998,Walgraef1997,Cross2009,Cross1993,Murray2002,Meinhardt1982,Maini2001}. For simplicity we omit, in this section, indices labeling lattice sites, and denote the receptor and scaffold numbers by $N^{r,s}$, and the corresponding mean-field receptor and scaffold concentrations by $r(t)$ and $s(t)$, respectively.

Previous studies \cite{Samoilov2006,Erban2009,Gillespie1976,Gillespie1977,Gillespie2013,McKane2004,Lugo2008,Kahraman2016} have shown that, for finite molecule numbers, the deterministic mean-field descriptions used in standard models of chemical dynamics \cite{Epstein1998,Walgraef1997,Cross2009,Cross1993,Murray2002,Meinhardt1982,Maini2001}
can fail to capture the average dynamics, as well as steady states, of the underlying stochastic reaction processes. As illustrated below, upper limits on the protein copy number due to molecular crowding, absorbing (non-fluctuating) states \cite{Samoilov2006}, bistability \cite{Samoilov2006}, and amplification of noise through nonlinear chemical reactions \cite{Erban2009} provide specific physical mechanisms yielding disagreement between MEs and mean-field equations. In Secs.~\ref{secSingleReact}--\ref{secFullReact} we consider reaction processes among receptors or scaffolds of increasing complexity, inspired by the reaction dynamics at synaptic domains (see Sec.~\ref{secModel}). We show that it is, at least in some special cases, practical to directly solve, either analytically or numerically, the MEs describing the chemical reaction dynamics at synaptic domains. We test the validity of these direct solutions of the MEs using KMC simulations, and provide systematic comparisons with solutions of the corresponding mean-field equations describing the reaction kinetics at synaptic domains.

\subsection{Single chemical reactions}
\label{secSingleReact}

\begin{figure*}[t!]
\includegraphics[width=2.1\columnwidth]{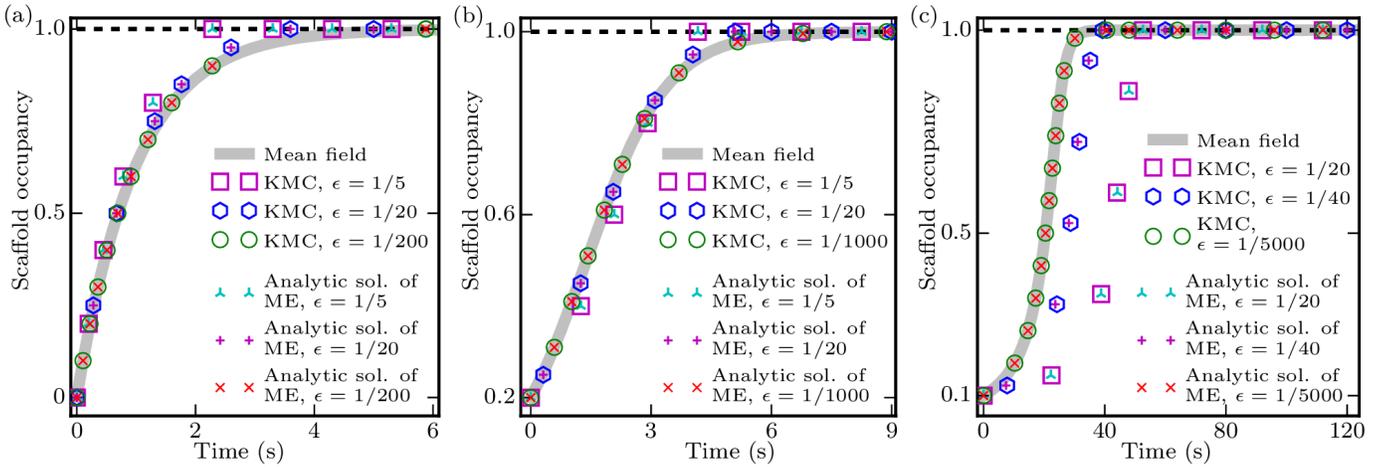} 
\caption{Average scaffold occupancies for (a) $S_b\xrightarrow{k_7}S$ with $N^s_0=0$, (b) $S_b+S\xrightarrow{\bar k_8}2S$ with $N^s_0=0.2$, and (c) $S_b+2S\xrightarrow{k_9}3S$ with $N^s_0=0.1$ obtained from the mean-field equations~(\ref{secReactMF1})--(\ref{secReactMF3}), the general analytic solutions of the MEs~(\ref{secReactME1})--(\ref{secReactME3}) for the mean jump times in Eq.~(\ref{eq:avg_jump_sol2MT}) with Eqs.~(\ref{eq:McEQ1})--(\ref{eq:McEQ3}),
and KMC simulations of the MEs~(\ref{secReactME1})--(\ref{secReactME3}).
We set $k_7=\bar k_8=k_9=1$~s$^{-1}$. The KMC simulations were averaged over 10$^5$ independent realizations each.}
\label{fig:single_reac}
\end{figure*}

In this section we focus on the single---linear and nonlinear---chemical reactions $S_b \xrightarrow{k_7} S$, $S_b +S \xrightarrow{\bar k_8} 2 S$, and $S_b + 2S \xrightarrow{k_9} 3S$ considered in Sec.~\ref{secModel}. We investigate each one of these reactions in turn. The reaction $S_b +S \xrightarrow{\bar k_8} 2 S$ is thereby implicit \cite{Haselwandter2015} in the reaction $M_b +S \xrightarrow{k_8} M_b + S_b$ discussed in Sec.~\ref{secModel}. Following Sec.~\ref{secModel}, the mean-field equations associated with these reactions are given by
\begin{align}
 \frac{ds}{dt} &= k_7 (1-s) \, , \label{secReactMF1} \\
 \frac{ds}{dt} &= \bar k_8 (1-s) s \, , \label{secReactMF2}\\
 \frac{ds}{dt} &= k_9 (1-s) \frac{s^2}{2} \, , \label{secReactMF3}
\end{align}
with the underlying MEs
{\allowdisplaybreaks\begin{align}
 \frac{d P(N^s,t)}{dt} =& -k_7 \frac{1-N^s}{\epsilon} P(N^s,t) \nonumber \\ \label{secReactME1}
 & + k_7 \frac{1-N^s+\epsilon}{\epsilon} P(N^s-\epsilon,t) \, ,\\
 \frac{d P(N^s,t)}{dt} =& -\bar k_8 \frac{1-N^s}{\epsilon} N^s P(N^s,t) \nonumber \\ \label{secReactME2}
 & + \bar k_8 \frac{1-N^s+\epsilon}{\epsilon} (N^s-\epsilon) P(N^s-\epsilon,t) \, , \\
 \frac{d P(N^s,t)}{dt} =& -k_9 \frac{1-N^s}{\epsilon} \frac{N^s(N^s-\epsilon)}{2} P(N^s,t) \nonumber \\
 & + k_9 \frac{1-N^s+\epsilon}{\epsilon} \frac{(N^s-\epsilon)(N^s-2\epsilon)}{2} \nonumber \\&\times P(N^s-\epsilon,t) \, . \label{secReactME3}
\end{align}}\noindent
We take the initial scaffold concentration at $t=0$ to be given by $N^s_0=s_0$
for the MEs and the mean-field equations, which then uniquely specifies the solutions of Eqs.~(\ref{secReactMF1})--(\ref{secReactMF3}) and Eqs.~(\ref{secReactME1})--(\ref{secReactME3}),
respectively. Note that Eqs.~(\ref{secReactME1})--(\ref{secReactME3}) exhibit
an absorbing state at $N^s=1$ with $P=1$ for $N^s=1$ and $P=0$ otherwise,
with Eqs.~(\ref{secReactME2}) and~(\ref{secReactME3}) also exhibiting an
absorbing state at $N^s=0$, and a further absorbing state at $N^s=\epsilon$
in Eq.~(\ref{secReactME3}). All transition rates that would allow the system to exit an absorbing state are, by definition, equal to zero. Fluctuations are therefore completely suppressed in an absorbing state, independent of the value of $\epsilon$ considered. The steady states of the mean-field equations~(\ref{secReactMF1})--(\ref{secReactMF3}) are given by $s=0$ or $s=1$.

The MEs~(\ref{secReactME1})--(\ref{secReactME3}) can be solved analytically, as follows. We first note that Eqs.~(\ref{secReactME1})--(\ref{secReactME3}) correspond to chains of Markov processes that irreversibly transform a system with $N_0^s/\epsilon$ scaffold molecules into a system with $1/\epsilon$ scaffold molecules. The longest possible chain of reactions is obtained with
$N_0^s=0$, for which
\begin{align} \label{eq:Markov_chain}
N^s=0 \xrightarrow{\alpha_0} N^s=\epsilon\xrightarrow{\alpha_1} ... \xrightarrow{\alpha_{C-1}} N^s=1\,,
\end{align}
where $C=1/\epsilon$ and the rates $\alpha_i$, in which $i={0,1,\dots,C-1}$, are
obtained from the first terms on the right-hand sides of the MEs~(\ref{secReactME1})--(\ref{secReactME3}):
\begin{align}  
\alpha_i =& k_7 \frac{1-i \epsilon}{\epsilon}  \,, \label{eq:McEQ1} \\ 
\alpha_i =& \bar k_8 \frac{(1-i \epsilon)}{\epsilon} i \epsilon\,, \label{eq:McEQ2} \\
\alpha_i =& k_9 \frac{(1-i \epsilon)}{\epsilon}
\frac{i \epsilon (i\epsilon-\epsilon)}{2} \,. \label{eq:McEQ3}
\end{align}
The probability distribution of jump times $t$ between two consecutive states $i$ and $i+1$ of the Markov processes in Eq.~(\ref{eq:Markov_chain}) is given
by the distribution of waiting times of a Poisson 
process with rate $\alpha_i$, 
\begin{align} \label{eq:EXP_dist}
 P_{i\rightarrow i+1}(t) = \alpha_i e^{-\alpha_i t} \, .
\end{align}
Equation (\ref{eq:EXP_dist}) with Eqs.~(\ref{eq:McEQ1})--(\ref{eq:McEQ3}) allows direct calculation of the probability distribution of jump times between arbitrary states in Eq.~(\ref{eq:Markov_chain}) (see Appendix~\ref{Appendix A}). However, to calculate the mean jump time between two arbitrary states
in Eq.~(\ref{eq:Markov_chain}) it is sufficient to note from Eq.~(\ref{eq:EXP_dist}) that the mean jump time from state $i$ to state $i+1$ is given by $\langle t \rangle_{i\to i+1}=1/\alpha_i$, implying a mean time
\begin{align} \label{eq:avg_jump_sol2MT}
\langle t \rangle_{p\to q} = \sum\limits_{i=p}^{q-1} \frac{1}{\alpha_i}
\end{align}
to reach a state $q$ with scaffold occupancy $N^s=q \epsilon$ starting from
an initial state $p=N^s_0/\epsilon$ (see Appendix~\ref{Appendix B}).

We find that the mean jump times predicted by Eq.~(\ref{eq:avg_jump_sol2MT}) with Eqs.~(\ref{eq:McEQ1})--(\ref{eq:McEQ3}) are in quantitative agreement with averages over KMC simulations of the corresponding MEs~(\ref{secReactME1})--(\ref{secReactME3}),
for all values of $\epsilon$ considered here (see Fig.~\ref{fig:single_reac}). To quantify the extent to which stochastic and mean-field results are
in agreement with each other in Fig.~\ref{fig:single_reac} we calculate, for each time point available from the MEs~(\ref{secReactME1})--(\ref{secReactME3}) in Fig.~\ref{fig:single_reac}, the difference between average stochastic and deterministic results, divide this difference by the corresponding mean-field result, and average the resultant percentage difference between stochastic and mean-field results over all time points in Fig.~\ref{fig:single_reac}
associated with a given reaction and value of $\epsilon$. For $\epsilon\gtrapprox1/5$ we find discrepancies $\gtrapprox10\%$ between the mean-field equations~(\ref{secReactMF1})--(\ref{secReactMF3}) and the MEs~(\ref{secReactME1})--(\ref{secReactME3}). The disagreement between
Eqs.~(\ref{secReactMF1})--(\ref{secReactMF3}) and Eqs.~(\ref{secReactME1})--(\ref{secReactME3})
becomes increasingly pronounced with increasing order of the reaction. For
instance, for the linear reaction $S_b \xrightarrow{k_7} S$, $\epsilon \lessapprox
1/200$ gives a discrepancy $\lessapprox0.5\%$ between the mean-field and master equations, while the reaction $S_b +2S\xrightarrow{k_9} 3S$ requires $\epsilon \lessapprox 1/5000$ for mean-field and master equations to be in
similarly good agreement. In particular, for the linear reaction $S_b \xrightarrow{k_7} S$ the ME~(\ref{secReactME1}) yields, for large enough $\epsilon$, a more rapid temporal evolution of the system than the corresponding mean-field equation~(\ref{secReactMF1}) [see Fig.~\ref{fig:single_reac}(a)]. This can be understood by noting that the ME allows for discrete jump processes with the system reaching the state $N^s=1$, on average, in a finite amount of time given by Eq.~(\ref{eq:avg_jump_sol2MT}), while the mean-field solution only reaches $s=1$ as $t \to \infty$. The same argument holds true for the nonlinear reactions in Fig.~\ref{fig:single_reac} [see Figs.~\ref{fig:single_reac}(b) and~\ref{fig:single_reac}(c)]. However,
for these nonlinear reactions we also find that, for a substantial portion of the trajectory of the system,
the mean jump time in the  MEs~(\ref{secReactME2}) and~(\ref{secReactME3})
is \textit{increased} compared to the corresponding mean-field results implied by Eqs.~(\ref{secReactMF2}) and~(\ref{secReactMF3}). These results illustrate that, already for very simple reaction dynamics, the molecular noise inherent in scaffold (and receptor) reaction dynamics can have subtle effects on average system properties if steric effects prohibit maximum molecule occupancies greater than hundreds of molecules, as is often the case for proteins in cell membranes.

\subsection{Competing chemical reactions}

In this section we study minimal reaction dynamics with competing chemical reactions increasing and decreasing the molecule number in the system. As model systems we use some of the key scaffold reactions at synaptic domains (see Sec.~\ref{secModel}). Reaction schemes with competing chemical reactions may, on the one hand, exhibit absorbing states in which all fluctuations are suppressed. On the other hand, competing chemical reactions may also yield fluctuating steady states of the ME with a mean that, for $\epsilon > 0$, does not necessarily coincide with the steady state implied by the corresponding mean-field model. We first consider reaction schemes for which the mean-field model predicts steady state(s) with $0<s<1$, while the ME yields absorbing state(s) at $N^s=0$ or $N^s=1$. Consistent with previous work \cite{Samoilov2006} we find that absorbing states can yield breakdown of the mean-field approach. We then consider reaction schemes with competing chemical reactions for which the MEs exhibit fluctuating steady states and no absorbing states. We find quantitative agreement between mean-field predictions and averages over the underlying MEs for linear reaction schemes, independent of the value of $\epsilon$ considered. However, we also find
that the mean-field approach can break down, for $\epsilon >0$, even for very simple nonlinear reaction schemes exhibiting fluctuating steady states.

\subsubsection{Absorbing states}

Consider a system composed of scaffolds undergoing the reactions $S\xrightarrow{k_6}S_b$ and $S_b+S\xrightarrow{k_9}2S$ discussed in Sec.~\ref{secModel}. Following
Sec.~\ref{secModel}, the corresponding mean-field equation is given by
\begin{equation} \label{secCompete1MF}
\frac{d s}{dt}=-k_6 s+k_9 (1-s) s\,, 
\end{equation}
with the underlying ME
\begin{align}
 \frac{d P(N^s,t)}{dt} =& -\frac{k_6}{\epsilon} \left[N^s P(N^s,t) \nonumber
  - \left(N^s+\epsilon\right) P(N^s+\epsilon,t) \right] \\
 & - \frac{k_9}{\epsilon} [\left(1-N^s\right) N^s P(N^s,t) \nonumber \\
 & + \left(1-N^s+\epsilon\right) (N^s-\epsilon) P(N^s-\epsilon,t)] \, .\label{secCompete1ME}
\end{align}
Equation~(\ref{secCompete1MF}) yields, for $k_9>k_6$, a stable steady state at $s=1-k_6/k_9$ and an unstable steady state at $s=0$, while Eq.~(\ref{secCompete1ME}) implies an absorbing state at $N^s=0$ with $P=1$ for $N^s=0$ and $P=0$ otherwise.

\begin{figure}[t!]
\includegraphics[width=\columnwidth]{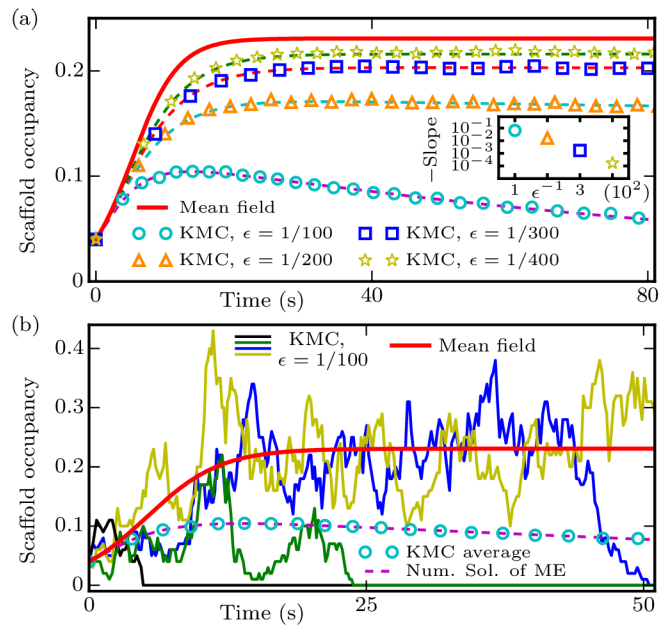} 
\caption{(a) Average scaffold occupancies for the reaction scheme $S\xrightarrow{k_6}S_b$ and $S_b+S\xrightarrow{k_9}2S$ obtained from the mean-field equation~(\ref{secCompete1MF})
and KMC simulations of the ME~(\ref{secCompete1ME}). The inset shows the average negative slopes of the average scaffold occupancies obtained
from the ME~(\ref{secCompete1ME}) for $\epsilon=1/400$, 1/300, 1/200, and 1/100, estimated from KMC data starting at the global maxima of the scaffold occupancies. The KMC simulations were averaged over 2000 independent realizations each. (b) Selected individual KMC trajectories for $\epsilon=1/100$ and corresponding mean-field solution as in panel (a). We set $k_6=1$~s$^{-1}$ and $k_9=1.3$~s$^{-1}$, and used an initial scaffold occupancy $N^s_0=0.04$. 
}
\label{fig:absorb}
\end{figure}

Direct solution of Eq.~(\ref{secCompete1MF}) for $k_9 > k_6$ confirms that, provided $s>0$ initially, the mean-field solution approaches the stable steady state $s=1-k_6/k_9$ as $t\to \infty$ [see Fig.~\ref{fig:absorb}(a)]. KMC simulations of Eq.~(\ref{secCompete1ME}) show that, initially, the average $N^s$ also tends to approach the stable steady state of the mean-field model [Fig.~\ref{fig:absorb}(a)]. Over time, however, fluctuations gradually carry the average of the stochastic system, with $\epsilon>0$, away from the steady state of the mean-field system and towards the absorbing state. This can be seen quite clearly by following individual stochastic trajectories of the system, which can get irreversibly locked into the absorbing state $N^s=0$ [see Fig.~\ref{fig:absorb}(b)]. Thus, the asymptotic behavior of the mean-field equation~(\ref{secCompete1MF}) is dominated by its stable steady state at $s=1-k_6/k_9$, while the asymptotic behavior of the underlying stochastic system is dominated by the absorbing state at $N^s=0$. As $\epsilon$ is increased, averages over the ME~(\ref{secCompete1ME}) approach the absorbing state increasingly rapidly, with the magnitude of the average slope increasing exponentially with decreasing maximum molecule occupancy (decreasing $1/\epsilon$) [see inset in Fig.~\ref{fig:absorb}(a)]. We find that, depending on the value of $\epsilon$ considered, it can take a substantial amount of time until the absorbing state is reached in our KMC simulations of the ME~(\ref{secCompete1ME}). For instance, with $\epsilon=1/400$ and $k_6$ and $k_9$ both being of the order of $1$~s$^{-1}$, only $\approx 3\%$ of the 2000 KMC trajectories in Fig.~\ref{fig:absorb}(a) are absorbed by $N^s=0$ for $t \lessapprox 4000$~s.

As a second example of a system with competing chemical reactions exhibiting
an absorbing state, consider the reactions $M_b+S\xrightarrow{k_8}M_b+S_b$ and $S_b+S\xrightarrow{\bar k_8}2S$ (see also Sec.~\ref{secSingleReact}). Following Sec.~\ref{secModel}, the corresponding mean-field equation is given by
\begin{align} \label{eq:mf_absorb2}
 \frac{ds}{dt} = -k_8(1-s)s + \bar k_8(1-s)s \,,
\end{align}
with the ME
\begin{align} \label{eq:ME_absorb2}
 \frac{dP}{dt} =& -\frac{k_8}{\epsilon} \big[ (1-N^s)N^s P(N^s,t) \nonumber \\ 
               & -(1-N^s-\epsilon)(N^s+\epsilon) P(N^s+\epsilon,t) \big] \nonumber \\
               & -\frac{\bar k_8}{\epsilon} \big[ (1-N^s)N^s P(N^s,t) \nonumber \\
               & -(1-N^s+\epsilon)(N^s-\epsilon) P(N^s-\epsilon,t) \big] \, .
\end{align}
The ME~(\ref{eq:ME_absorb2}) has two absorbing states, at $N^s=0$ and $N^s=1$.
For $k_8 \neq \bar k_8$, the mean-field equation~(\ref{eq:mf_absorb2}) exhibits steady states at $s=0$ and $s=1$, with the steady state $s=0$ ($s=1$) being unstable (stable) for $k_8 < \bar k_8$, and vice versa for $k_8 > \bar k_8$. For $k_8 = \bar k_8$, the right-hand side of the mean-field equation~(\ref{eq:mf_absorb2}) is identical to zero, with any initial condition $s(0)=s_0$ corresponding to a steady state of the system.

\begin{figure}[t!]
\includegraphics[width=\columnwidth]{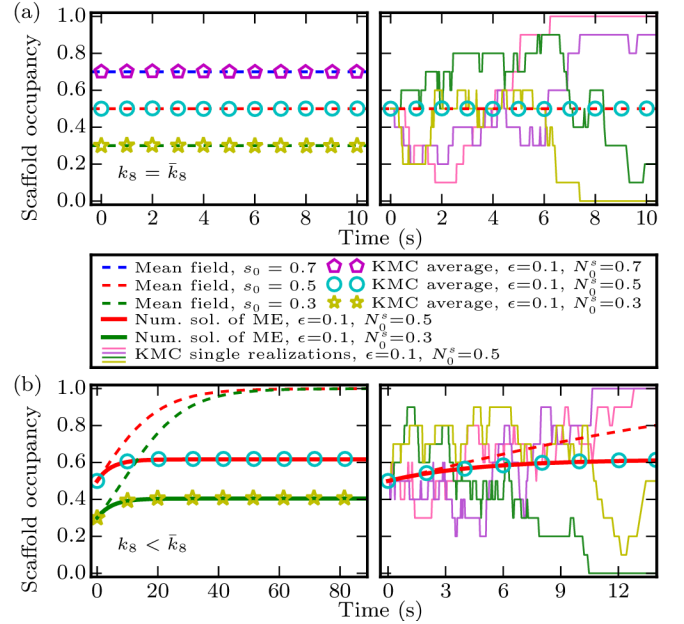} 
\caption{Scaffold occupancies for the reaction scheme $M_b+S\xrightarrow{k_8}M_b+S_b$ and $S_b+S\xrightarrow{\bar k_8}2S$ obtained from the mean-field equation~(\ref{eq:mf_absorb2})
and KMC simulations or direct numerical solutions of the ME~(\ref{eq:ME_absorb2})
for (a) $k_8=\bar k_8=1.0$~s$^{-1}$ and (b) $k_8=1.0$~s$^{-1}$ and $\bar k_8=1.1$~s$^{-1}$. The left panels show mean scaffold occupancies, with the
KMC simulations averaged over $2\times10^4$ independent realizations each, and the right panels show selected individual KMC trajectories of the system,
together with the corresponding mean-field results. We use the same labeling conventions in panel (b) as in panel (a).
}
\label{fig:absorb2}
\end{figure}

Comparing KMC simulations of the ME~(\ref{eq:ME_absorb2}) with solutions of the mean-field equation~(\ref{eq:mf_absorb2}) we find that, for $k_8 = \bar k_8$, the mean-field equation~(\ref{eq:mf_absorb2}) is in quantitative agreement with averages over the ME~(\ref{eq:ME_absorb2}), independent of the initial conditions used [see the left panel of Fig.~\ref{fig:absorb2}(a)]. However, the average scaffold concentration fails to capture the asymptotic properties of individual KMC trajectories which, for long enough times, are absorbed by either $N^s=0$ or $N^s=1$, and hence do not fluctuate
about the average scaffold concentration [see the right panel of Fig.~\ref{fig:absorb2}(a)].
For $k_8 \neq \bar k_8$, we find that the mean-field equation~(\ref{eq:mf_absorb2})
fails to capture averages over the ME~(\ref{eq:ME_absorb2}) with $\epsilon>0$
obtained via direct numerical solution of the ME~(\ref{eq:ME_absorb2}) and KMC simulations [see the left panel of Fig.~\ref{fig:absorb2}(b)]. Similarly as for $k_8 = \bar k_8$, the properties of individual stochastic trajectories of the system are dominated, at long times, by the absorbing states $N^s=0$ or $N^s=1$ rather than the average scaffold concentration [see the right panel of Fig.~\ref{fig:absorb2}(b)]. We have confirmed the results of our
KMC simulations and our numerical solutions of the ME~(\ref{eq:ME_absorb2})
through analytic solution of the ME~(\ref{eq:ME_absorb2}) at steady state.
Our results illustrate \cite{Samoilov2006} how, for $\epsilon>0$, absorbing states can yield breakdown of the mean-field reaction dynamics at synaptic domains, and produce pronounced discrepancies between individual stochastic trajectories of the system and ensemble averages.

\subsubsection{Fluctuating steady states}

In this section we focus on linear and nonlinear scaffold reaction dynamics with no absorbing states but (fluctuating) steady states. We first consider the linear reactions $S \xrightarrow{k_6} S_b$ and $S_b \xrightarrow{k_7} S$ discussed in Sec.~\ref{secModel}, yielding the mean-field equation
\begin{align} \label{eq:mf_dual1}
 \frac{d s}{dt} = -k_6 s + k_7 (1-s) \,,
\end{align}
with the stable steady state $s=k_7/\left(k_6+k_7\right)$, and the~ME
\begin{align} \label{eq:ME_dual1}
 \frac{d P(N^s,t)}{dt} =& -\frac{k_6}{\epsilon} \left[N^s P(N^s,t) \nonumber
  - \left(N^s+\epsilon\right) P(N^s+\epsilon,t) \right]
  \nonumber\\
 & +\frac{k_7}{\epsilon} \bigl[\left(1-N^s+\epsilon\right) P\left(N^s-\epsilon,t\right) \nonumber\\
 & - \left(1-N^s\right) P\left(N^s,t\right) \bigr] \,.
\end{align}

To analytically determine the steady-state probability distribution(s) of the ME~(\ref{eq:ME_dual1}), $P_\infty(N^s)$, we generalize the generating-function approach for the solution of MEs described in Ref.~\cite{Gardiner1985} to allow for steric constraints. To this end, we define a generating function
\begin{align} \label{eq:generating_fun}
G(\hat s) = \sum_{n=0}^{1/\epsilon} {\hat s}^n P_\infty\left(N^s\right) \, ,
\end{align}
where $N^s=n \epsilon$ and we only include terms corresponding to $0\leq N^s \leq 1$ because, by definition, $P_\infty(N^s)=0$ outside this range. The generating function in Eq.~(\ref{eq:generating_fun}) satisfies the following identities:
\begin{eqnarray} \label{eq:GF_relation1}
 \hat s \frac{d G(\hat s)}{d \hat s} &=& \sum_{n=0}^{1/\epsilon} n \hat s^n P_\infty\left(n\epsilon\right) \, ,  \\ \label{eq:GF_relation2}
  \frac{d G(\hat s)}{d \hat s} &=& \sum_{n=0}^{1/\epsilon} (n+1) \hat s^n P_\infty\left((n+1)\epsilon\right) \, .
\end{eqnarray}
It then follows that
\begin{equation} \label{eq:GF_relation3}
\frac{\hat s}{\epsilon}G(\hat s) - \hat s^2 \frac{d G(\hat s)}{d \hat s} = \sum_{n=0}^{1/\epsilon} \left(\frac{1}{\epsilon}-n+1\right) \hat s^n P_\infty\left((n-1)\epsilon\right)\,,
\end{equation}
where we have used that, by definition, $P_\infty(N^s)=0$ for $N^s<0$. Setting the left-hand side of the ME~(\ref{eq:ME_dual1}) equal to zero, multiplying the right-hand side by ${\hat s}^n$, summing all terms from $n=0$ to $n=1/\epsilon$, and employing the identities in Eqs.~(\ref{eq:GF_relation1})--(\ref{eq:GF_relation3}), we find that, in the steady state(s) of Eq.~(\ref{eq:ME_dual1}), the generating function obeys
\begin{align}\label{secReactGdiffEq}
\frac{d G(\hat s)}{d \hat s} = \frac{1}{\epsilon} \frac{k_7}{k_6 + k_7 \hat s } G\,.
\end{align}

\begin{figure}[t!]
\includegraphics[width=\columnwidth]{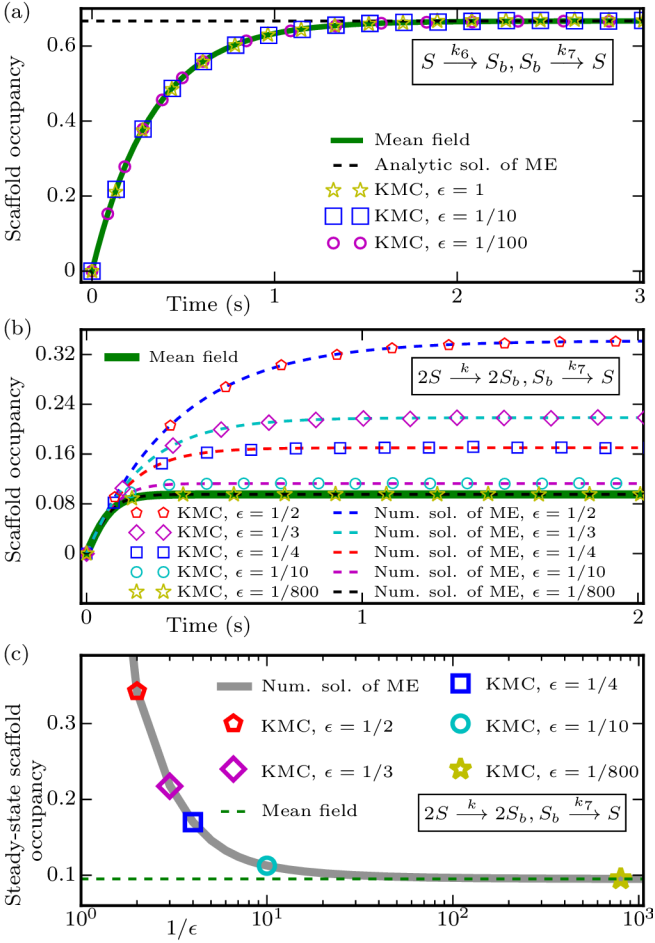} 
\caption{
Fluctuating steady states. Average scaffold occupancies for (a) $S \xrightarrow{k_6} S_b$ and $S_b \xrightarrow{k_7} S$ obtained from the
mean-field equation~(\ref{eq:mf_dual1}), the steady-state distribution of the ME~(\ref{eq:ME_dual1}) in Eq.~(\ref{eq:binomail}), and KMC simulations of the ME~(\ref{eq:ME_dual1}), and (b,c) $2S\xrightarrow{k}2S_b$ and $S_b\xrightarrow{k_7}S$ obtained from the mean-field equation~(\ref{eq:mf_dual2}), direct numerical
(steady-state) solutions of the ME~(\ref{eq:ME_dual2}), and KMC simulations of the ME~(\ref{eq:ME_dual2}). In panel (c), the steady-state scaffold occupancy in Eq.~(\ref{eq:mf_sol_dual2}) implied by the mean-field equation~(\ref{eq:mf_dual2}), $s\approx 0.095$, is indicated by a horizontal dashed line. For panel (a) we set $k_6=1$~s$^{-1}$ and $k_7=2$~s$^{-1}$, and for panels (b) and (c) we set $k=100$~s$^{-1}$ and $k_7=1$~s$^{-1}$. For all panels we used an initial scaffold occupancy $N^s_0=s_0=0$. All KMC simulations were averaged over $2\times10^4$ independent realizations each.
}
\label{fig:dual_reac}
\end{figure}

Equation~(\ref{secReactGdiffEq}) has the unique solution
\begin{align} \label{secreactGsolB}
G(\hat s)= \left(\frac{k_6 +k_7 \hat s}{k_6+k_7}\right) ^{\frac{1}{\epsilon}}
\end{align}
satisfying the normalization constraint $G(1)=1$. Hence, Eq.~(\ref{eq:ME_dual1}) admits only one steady-state probability distribution. To determine the form of this distribution we note from Eq.~(\ref{eq:generating_fun}) that
\begin{equation} \label{secreactGsolBsub}
P_\infty(n\epsilon) = \frac{1}{n!}\frac{d^n G(\hat s)}{d \hat s^n}\bigg|_{\hat s=0}. 
\end{equation}
Equations~(\ref{secreactGsolB}) and~(\ref{secreactGsolBsub}) imply that the steady-state probability distribution associated with Eq.~(\ref{eq:ME_dual1})
takes the form of a binomial distribution,
\begin{eqnarray}
P_\infty(N^s) &=& \frac{\frac{1}{\epsilon}!}{n!\left(\frac{1}{\epsilon}-n\right)!}
\left(\frac{k_7}{k_6+k_7}\right)^{n}
  \nonumber \\ && \left(1-\frac{k_7}{k_6+k_7}\right)^{\frac{1}{\epsilon}-n}\,,
  \label{eq:binomail}
\end{eqnarray}
where $n=N^s/\epsilon$. Equation~(\ref{eq:binomail}) shows that the mean scaffold occupancy at steady state is given by $\langle N_s \rangle=k_7/\left(k_6+k_7\right)$, which is identical to the steady-state value of $s$ predicted by the mean-field equation~(\ref{eq:mf_dual1}) independent of the value of $\epsilon$ in Eq.~(\ref{eq:ME_dual1}). As exemplified by Eq.~(\ref{eq:binomail}), fluctuating steady states allow, in contrast to absorbing states, fluctuations about the mean molecule occupancy. As expected from the above considerations, solution of the mean-field equation~(\ref{eq:mf_dual1}),
the average $N^s$ implied by Eq.~(\ref{eq:binomail}), and averages over KMC simulations of the ME~(\ref{eq:ME_dual1}) yield, even for $\epsilon=1$, identical results for the steady state of the system [see Fig.~\ref{fig:dual_reac}(a)]. We also find that the mean-field model yields quantitative agreement with KMC simulations of the ME~(\ref{eq:ME_dual1}) for transient regimes of the system, independent of the value of $\epsilon$
considered [Fig.~\ref{fig:dual_reac}(a)].

The two most straightforward nonlinear versions of the reaction scheme
considered above are $S+M_b \xrightarrow{k_8} S_b$ and $S_b \xrightarrow{k_7} S$, and $S \xrightarrow{k_6} S_b$ and $S_b+S \xrightarrow{\bar k_8} 2 S$ (see Sec.~\ref{secModel}). However, both of these reaction schemes exhibit absorbing states, which makes them unsuitable for the purposes of the present discussion. Instead, to explore the properties of nonlinear fluctuating steady states in a simple model system, we consider the reactions $2S\xrightarrow{k}2S_b$ and $S_b\xrightarrow{k_7}S$. The first of these reactions may not be relevant for synaptic domains \cite{Haselwandter2015,Haselwandter2011}, but may occur in other contexts \cite{Erban2007,Walgraef1997,Cross2009,Cross1993,Murray2002}. Following similar steps as in Sec.~\ref{secModel}, we find that the mean-field equation associated with this reaction scheme is given by
\begin{align} \label{eq:mf_dual2}
  \frac{d s}{dt} = k_7 (1-s) -k s^2 \, ,
\end{align}
yielding a stable steady state at
\begin{align} \label{eq:mf_sol_dual2}
s = \frac{-k_7 + \sqrt{k_7(4 k +k_7)}}{2k} \, ,
\end{align}
with the second steady state of Eq.~(\ref{eq:mf_dual2}) lying outside the
physically relevant range $0\leq s \leq 1$. The underlying ME is given by
\begin{align} \label{eq:ME_dual2}
  \frac{d P(N^s,t)}{dt} =& \frac{k_7}{\epsilon} \bigl[\left(1-N^s-\epsilon\right) P\left(N^s-\epsilon,t\right) \nonumber\\
 & - \left(1-N^s\right) P\left(N^s,t\right) \bigr]\nonumber\\
  &-\frac{k}{2! \epsilon} \bigl[N^s \left(N^s-\epsilon\right)
  P(N^s,t) \nonumber\\
  & - \left(N^s+2\epsilon\right) \left(N^s+\epsilon\right) P(N^s+2\epsilon,t)\bigr] \,.
\end{align}  
We ascertain the properties of the ME~(\ref{eq:ME_dual2}) using direct numerical solutions of Eq.~(\ref{eq:ME_dual2}) as well as KMC simulations. We find quantitative agreement between direct numerical solutions of the ME~(\ref{eq:ME_dual2}) and KMC simulations for all values of $\epsilon$ considered here [see Fig.~\ref{fig:dual_reac}(b)]. For $\epsilon\gtrapprox 1/10$, however, the mean-field equation~(\ref{eq:mf_dual2}) fails to predict the average scaffold concentrations implied by the ME~(\ref{eq:ME_dual2}) [Fig.~\ref{fig:dual_reac}(b)]. As $\epsilon$ is decreased, averages over the steady-state scaffold occupancies implied by the ME~(\ref{eq:ME_dual2}) approach the steady-state value of $s$ in Eq.~(\ref{eq:mf_sol_dual2}),
with agreement between stochastic and mean-field results for $\epsilon \lessapprox 1/10$ [see Fig.~\ref{fig:dual_reac}(c)].

\subsection{Reaction kinetics at synaptic domains}
\label{secFullReact}

\begin{figure}[b!]
\includegraphics[width=\columnwidth]{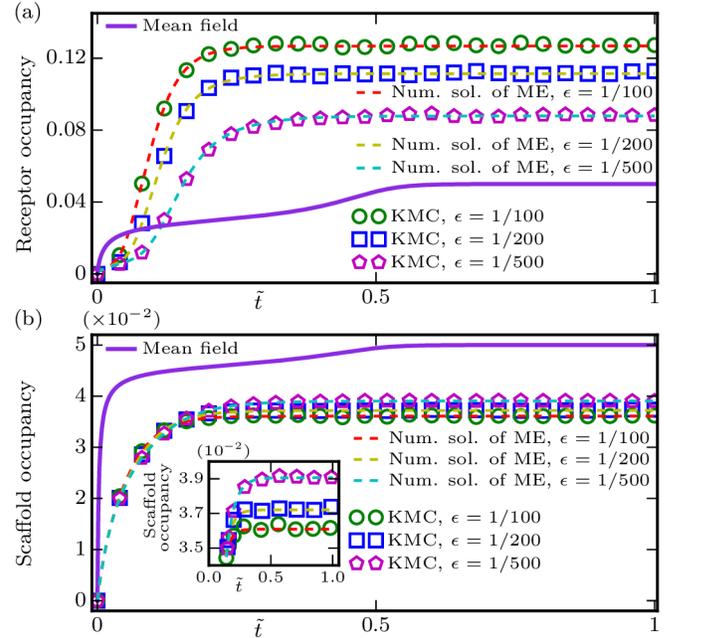} 
\caption{Average (a) receptor and (b) scaffold occupancies in a reaction-only
system with the reaction kinetics at synaptic domains described in Sec.~\ref{secModel}
(see Table~\ref{tabReactionRates}) \cite{Haselwandter2011,Haselwandter2015,Kahraman2016}, obtained from direct numerical solutions of the ME~(\ref{eq:ME}), KMC simulations of the ME~(\ref{eq:ME}), and the mean-field equations~(\ref{eq:MFE_r}) and~(\ref{eq:MFE_s}) versus scaled time $\tilde{t}=t/\tau$, with $\tau=1.0\times10^4$~s
and $\tau=5.0\times10^5$~s for the stochastic and mean-field models, respectively.
The inset in panel (b) shows how the average scaffold occupancies implied
by the ME~(\ref{eq:ME}) change with $\epsilon$. All KMC simulations were averaged over $2\times10^4$ independent realizations
each.}
\label{fig:full_reac}
\end{figure}

In this section we consider the complete model of the reaction kinetics
at synaptic domains described in Sec.~\ref{secModel}, which exhibits coupled, nonlinear reactions among receptors or scaffolds. We have shown previously \cite{Kahraman2016} that, for a maximum molecule occupancy per membrane patch $1/\epsilon \approx 100$, which is the regime of $\epsilon$ relevant for cell membranes (see also Sec.~\ref{secModel}), the mean-field equations~(\ref{eq:MFE_r}) and~(\ref{eq:MFE_s}) fail to capture the temporal evolution as well as steady-state values of the average receptor and scaffold concentrations implied by the ME~(\ref{eq:ME}), with the mean-field system approaching its steady state approximately one order of magnitude slower than the stochastic system. We have confirmed these conclusions using direct numerical solutions of the ME~(\ref{eq:ME}), which we find to be in quantitative agreement with KMC simulations of the ME~(\ref{eq:ME}) (see Fig.~\ref{fig:full_reac}).

\begin{figure}[t!]
\includegraphics[width=\columnwidth]{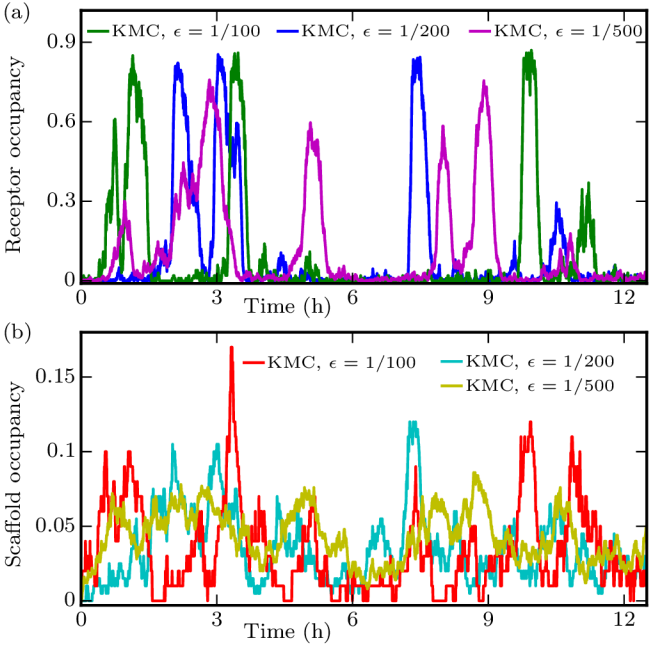} 
\caption{Individual stochastic trajectories of (a) receptor and (b) scaffold occupancies in a reaction-only system with the reaction kinetics at synaptic domains described in Sec.~\ref{secModel} (see Table~\ref{tabReactionRates}) \cite{Haselwandter2011,Haselwandter2015,Kahraman2016}, obtained from the KMC data in Fig.~\ref{fig:full_reac}. 
}
\label{fig:full_reac_b}
\end{figure}

\begin{figure}[t!]
\includegraphics[width=\columnwidth]{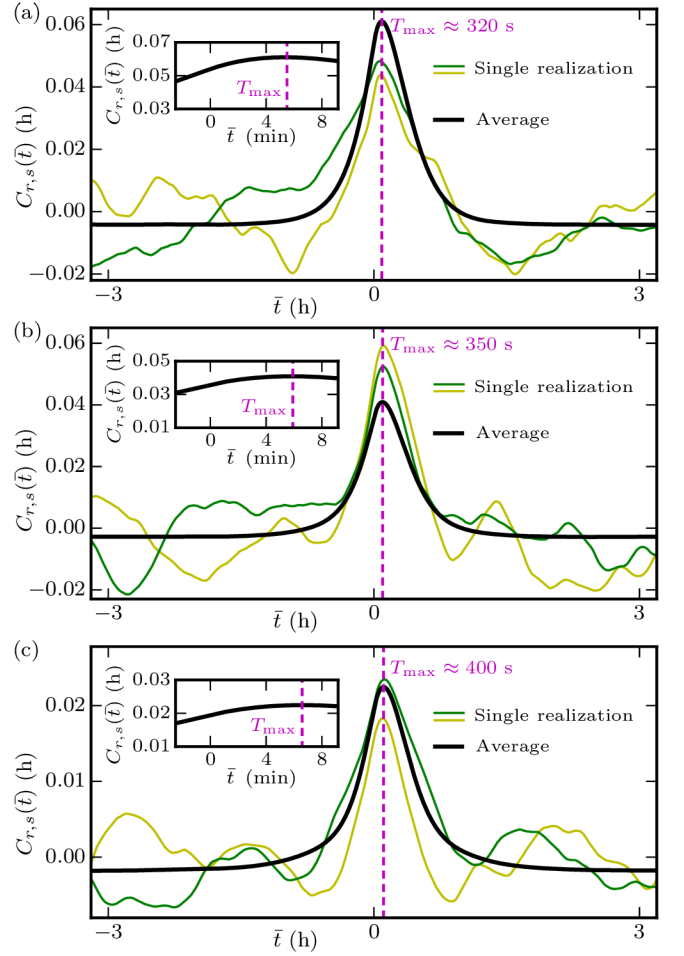} 
\caption{Receptor-scaffold correlation function in Eq.~(\ref{eq:correlation}) for (a) $\epsilon=1/100$, (b) $\epsilon=1/200$, and (c) $\epsilon=1/500$
in a reaction-only system with the reaction kinetics at synaptic domains described in Sec.~\ref{secModel} (see Table~\ref{tabReactionRates}) \cite{Haselwandter2011,Haselwandter2015,Kahraman2016}, obtained from KMC simulations of the ME~(\ref{eq:ME}) as in Fig.~\ref{fig:full_reac}.
The yellow and green curves show $C_{r,s}(\bar t)$ computed for single KMC
trajectories, while the black curves show the average $C_{r,s}(\bar t)$ obtained, as in Fig.~\ref{fig:full_reac}, from $2\times10^4$ independent realizations each. The vertical dashed lines in the main panels and insets indicate the locations of the global maxima of $C_{r,s}(\bar t)$ at $\bar t=T_\textrm{max}$.}
\label{fig:full_reac_c}
\end{figure}

To further explore the stochastic reaction dynamics at synaptic domains, we follow individual stochastic trajectories of the system (see Fig.~\ref{fig:full_reac_b}). Inspection of individual stochastic trajectories shows that, even in the steady state of the system, the receptor and scaffold occupancies can undergo large fluctuations, with the fluctuations in the receptor occupancy being particularly pronounced. To quantify the correlation between receptor and scaffold fluctuations, we calculate the receptor-scaffold correlation function in the steady-state of the stochastic system,
\begin{eqnarray} 
C_{r,s}(\bar t) &=& \int\limits_{t_\mathrm{min}}^{t_\mathrm{max}} dt \left[N^r(t)-\langle N^r \rangle\right] \left[N^s(t+\bar t)-\langle N^s \rangle\right] \, , \nonumber \\
 \label{eq:correlation}
\end{eqnarray}
in which we set $t_\mathrm{min}=1.0\times10^4$~s and $t_\mathrm{max}=1.3\times10^5$~s, $\langle N^{r,s}\rangle$ are the average receptor and scaffold occupancies in the time interval $[t_\mathrm{min},t_\mathrm{max}]$ implied by our KMC simulations, and we employ periodic boundary conditions when computing $N^{s}(t)$. We selected the value $t_\mathrm{min}=1.0\times10^4$~s in Eq.~(\ref{eq:correlation}) so that, for all the values of $\epsilon$ considered here, the average receptor and scaffold occupancies in Fig.~\ref{fig:full_reac} have reached their steady-state values at $t=t_\mathrm{min}$.

For all the values of $\epsilon$ considered here, we find that the time of maximum correlation in Eq.~(\ref{eq:correlation}) occurs for $\bar t >0$ (see Fig.~\ref{fig:full_reac_c}). This indicates that, consistent with the roles of receptors and scaffolds as ``inhibitors'' and ``activators'' of increased molecule occupancies \cite{Haselwandter2011,Haselwandter2015}, fluctuations increasing the scaffold occupancy trigger increased receptor occupancies. In turn, increased receptor occupancies tend to inhibit increased receptor as well as scaffold occupancies \cite{Haselwandter2011,Haselwandter2015}, with fluctuations decreasing the scaffold occupancy  precipitating decreased receptor occupancies. Taken together, Figs.~\ref{fig:full_reac_b} and~\ref{fig:full_reac_c} thus suggest that small fluctuations in the scaffold occupancy can trigger large changes in the receptor occupancy, with the feedback between receptor and scaffold occupancies effectively amplifying receptor fluctuations. Figure~\ref{fig:full_reac_c} also suggests that the time of maximum correlation between fluctuations in the receptor and scaffold occupancies grows with decreasing $\epsilon$. Finally, we note that calculation of the power spectrum \cite{Butler2009,Butler2011} of the fluctuating receptor and scaffold occupancies obtained from our KMC simulations does not yield a characteristic (non-zero) frequency of receptor and scaffold fluctuations.

\begin{figure}[t!]
\includegraphics[width=\columnwidth]{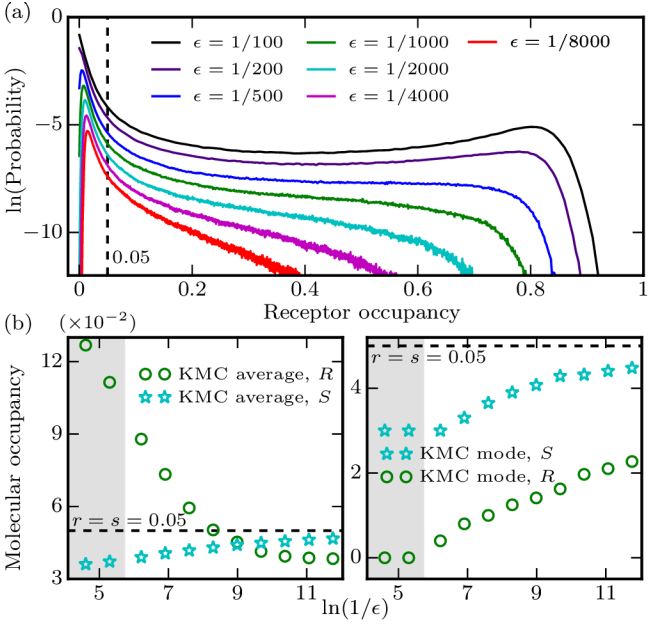} 
\caption{Statistical properties of the steady-state receptor and scaffold
occupancies implied by the stochastic reaction dynamics at synaptic domains
described in Sec.~\ref{secModel} (see Table~\ref{tabReactionRates}) \cite{Haselwandter2011,Haselwandter2015,Kahraman2016}.
(a) Marginal steady-state probability distributions of the receptor occupancy for the indicated values of $\epsilon$. (b) Average (left panel) and mode (right panel) of the steady-state receptor and scaffold occupancies versus $1/\epsilon$, obtained from the respective marginal steady-state probability distributions. The grey regions show the approximate range of $\epsilon$ for which the receptor and scaffold probability distributions are bistable. For such values of $\epsilon$, we use the global maximum as the mode. The steady-state values of $r$ and $s$ implied by the mean-field equations~(\ref{eq:MFE_r}) and~(\ref{eq:MFE_s}), $r=s=0.05$, are indicated by a vertical dashed line in panel (a) and by horizontal dashed lines in panel (b). All KMC simulations were averaged, from $t_\textrm{min}=1.0\times10^5$~s to $t_\textrm{max}=4.5\times10^5$~s,
over $10^4$ independent realizations each. 
} \label{fig:full_reac2}
\end{figure}

The KMC trajectories in Fig.~\ref{fig:full_reac_b} suggest that, for large enough $\epsilon$, the steady-state receptor occupancy is bistable, fluctuating between $N^r \gtrapprox 0.8$ and a value of $N^r$ close to zero. To further
quantify the fluctuations in the receptor occupancy, we used our KMC data
to estimate the steady-state probability distribution of the receptor occupancy, marginalized over the scaffold distribution, for different values of $\epsilon$ [see Fig.~\ref{fig:full_reac2}(a)]. Consistent with Fig.~\ref{fig:full_reac_b} we find that, for the value $\epsilon\approx1/100$ relevant for synaptic domains \cite{Kahraman2016}, the receptor occupancy is bistable, with maxima at $N^r \approx 0$ and $N^r \approx 0.8$. As $\epsilon$ is decreased below $\epsilon \approx 1/300$, we find a unique maximum (mode) of the probability distribution. However, even for $\epsilon \approx 7.8 \times 10^{-6}$, which would correspond to a ``well-mixed'' membrane compartment holding up to $\approx
1.28 \times 10^5$ receptors or scaffolds, we find that both the averages [see Fig.~\ref{fig:full_reac2}(b)] and modes [see Fig.~\ref{fig:full_reac2}(c)] of the steady-state receptor and scaffold probability distributions do not coincide with the steady-state receptor and scaffold occupancies implied by the mean-field equations~(\ref{eq:MFE_r}) and~(\ref{eq:MFE_s}), with the disagreement being more pronounced for receptors than scaffolds.

\begin{figure}[t!]
\centering
\includegraphics[width=\columnwidth]{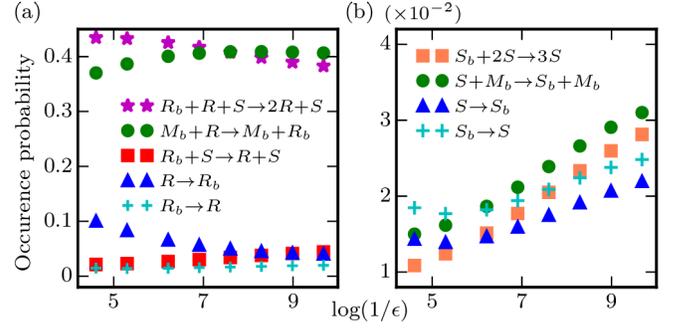}
\caption{Occurrence probabilities of individual chemical reactions for the reaction dynamics at synaptic domains in Sec.~\ref{secModel} (see Table~\ref{tabReactionRates}) \cite{Haselwandter2011,Haselwandter2015,Kahraman2016} versus $1/\epsilon$. The occurrence probabilities were computed at steady state using KMC simulations of the ME~(\ref{eq:ME}) by calculating, for the time interval $[t_{\rm{min}}, t_{\rm{max}}]$ with $t_\textrm{min}=1.0\times10^5$~s and $t_\textrm{max}=2.0\times10^5$~s, the ratio of the occurrence number of a particular reaction and the total occurrence number of all the (receptor and scaffold) reactions in the system.
The KMC simulations were averaged over $10^4$ independent realizations.
}
\label{fig:reaction_vs_eps}
\end{figure}

The above results show that for, $\epsilon>0$, the ME~(\ref{eq:ME}) and the
corresponding mean-field equations~(\ref{eq:MFE_r}) and~(\ref{eq:MFE_s})
can yield qualitatively and quantitatively different results for the reaction kinetics at synaptic domains discussed in Sec.~\ref{secModel} \cite{Haselwandter2011,Haselwandter2015,Kahraman2016}.
We further characterize the dependence of the solutions of the ME~(\ref{eq:ME})
on $\epsilon$ by calculating the occurrence probabilities of the different chemical reactions comprising the reaction dynamics at synaptic domains (see Table~\ref{tabReactionRates}), as a function of $\epsilon$ (see Fig.~\ref{fig:reaction_vs_eps}). For the receptors we find that, as $\epsilon$ is decreased, the occurrence probabilities of the reactions $R\xrightarrow{k_1}R_b$ and $R_b+R+S\xrightarrow{k_5}2R+S$
tend to decrease compared to the occurrence probabilities of the other reactions
changing the receptor number in the system. For the scaffolds we find that, as $\epsilon$ is decreased, the occurrence probability of the reaction $S\xrightarrow{k_6}S_b$ tends to decrease compared to the occurrence probabilities of the other reactions changing the scaffold number in the system. As further discussed in Sec.~\ref{secRSD}, calculation of the occurrence probabilities of receptor and scaffold reactions across synaptic domains permits insights into possible physical mechanisms
underlying spatially heterogeneous reaction dynamics at synaptic domains \cite{Kahraman2016,Czondor2012,Blanpied2002,Earnshaw2006}.

\section{Stochastic reaction-diffusion dynamics at synaptic domains}
\label{secRSD}

In this section we consider the full reaction-diffusion dynamics at synaptic
domains (see Sec.~\ref{secModel}). We first study, and contrast, the collective properties of synaptic domains obtained from our stochastic and mean-field
models of receptor-scaffold reaction-diffusion dynamics at synaptic domains. We then use our stochastic lattice model to compute the occurrence probabilities of receptor and scaffold reactions across synaptic domains. We find that the reaction-diffusion processes at synaptic domains give rise to characteristic locations of receptor and scaffold insertion and removal at synaptic
domains. Next, we use our stochastic lattice model to study global receptor and scaffold turnover at synaptic domains \cite{Kahraman2016}, and make comparisons to the results of FRAP experiments \cite{Choquet2003,Specht2008,Calamai2009}. Finally, we show \cite{Kahraman2016} that our stochastic lattice model of the reaction-diffusion dynamics at synaptic domains yields single-molecule trajectories consistent with experimental observations 
\cite{Choquet2003,Choquet2013,Meier2001,Borgdorff2002,Dahan2003,Specht2008,Triller2005,Triller2008},
and allows prediction of the time scale of receptor trafficking between membrane
regions inside and outside synaptic domains.

\subsection{Collective properties of synaptic domains}
\label{secSynColl}

\begin{figure}[t!]
\centering
\includegraphics[width=\columnwidth]{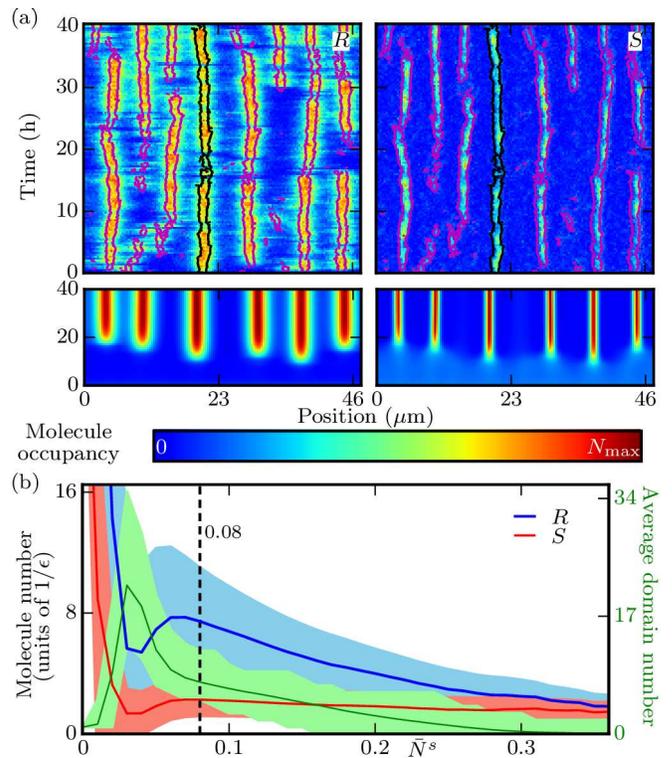} 
\caption{Self-assembly of synaptic domains. (a) Synaptic domains obtained from the ME~(\ref{eq:ME}) via KMC simulation (upper panels) and the mean-field equations~(\ref{eq:MFE_r}) and~(\ref{eq:MFE_s}) (lower panels) with the reaction-diffusion dynamics described in Sec.~\ref{secModel} using $\epsilon=1/100$. The left and right panels show the receptor and scaffold occupancies with maximum occupancies $(N_i^r,N_i^s)=(0.79,0.55)$ (KMC) and $(r,s)=(0.46,0.15)$ (mean field). The curves in the upper panels delineate the domain boundaries obtained using a threshold $\bar N^s=0.08$ on the scaffold occupancy of membrane patches. (b) Average receptor and scaffold numbers per synaptic domain and average number of domains in the system in the upper panels of (a), versus $\bar N^s$. The shaded blue (red) area shows the standard deviation of the receptor (scaffold) number per synaptic domain, while the shaded green area shows the minimum and maximum of the domain number. All data was extracted from the KMC results shown in the upper panels of (a).
}
\label{fig:patterns0}
\end{figure}

KMC simulations of the ME~(\ref{eq:ME}) show that the stochastic reaction-diffusion dynamics considered here (see Sec.~\ref{secModel}) yield \cite{Kahraman2016}, starting from random initial conditions, in-phase receptor and scaffold domains [see Fig.~\ref{fig:patterns0}(a)]. Using the same values of the reaction and diffusion rates as in the ME~(\ref{eq:ME}), the mean-field equations~(\ref{eq:MFE_r}) and~(\ref{eq:MFE_s}) yield self-assembly of stable receptor-scaffold domains of a similar characteristic wavelength $\approx8.5$~$\mu$m as found in our KMC simulations [Fig.~\ref{fig:patterns0}(a)], which is also consistent with the linear stability analysis of the mean-field equations~(\ref{eq:MFE_r}) and~(\ref{eq:MFE_s}) \cite{Haselwandter2011,Haselwandter2015}. Similarly as in 2D systems, where the reaction and diffusion rates used here yield \cite{Haselwandter2011,Haselwandter2015,Kahraman2016} synaptic domains of a similar characteristic size as observed in experiments 
\cite{Kirsch1995,Meier2000,Meier2001,Borgdorff2002,Dahan2003,Hanus2006,Ehrensperger2007,Calamai2009,Haselwandter2011}, the 1D patterns obtained from both the ME~(\ref{eq:ME}) and the mean-field equations~(\ref{eq:MFE_r}) and~(\ref{eq:MFE_s}) are irregular, with substantial variation in the spacing between individual domains. Furthermore, the ME~(\ref{eq:ME}) yields, for $\epsilon\approx1/100$, substantial fluctuations in the size and location of synaptic domains, over a time scale of several hours. Consistent with the results on the reaction-only system in Sec.~\ref{secFullReact}, we find that, for $\epsilon\approx1/100$, domain formation proceeds more rapidly in the stochastic lattice model than in the mean-field model \cite{Kahraman2016}, by approximately one order of magnitude. Thus, the molecular noise inherent in receptor and scaffold reaction and diffusion processes accelerates the self-assembly of synaptic~domains.

Consistent with our mean-field model \cite{Haselwandter2011,Haselwandter2015},
we find \cite{Kahraman2016} that the ME~(\ref{eq:ME}) yields scaffold profiles across synaptic domains that tend to be more narrow than receptor profiles [Fig.~\ref{fig:patterns0}(a)]. Scaffold domains are therefore more sharply defined than receptor domains, and it is convenient \cite{Kahraman2016} to specify domain boundaries in our stochastic lattice model by placing a threshold on the scaffold occupancy. We first remove small-scale fluctuations in the
scaffold occupancy using a Savitzky-Golay filter \cite{Savitzky1964} (order 5, frame size 25), and then apply a threshold $\bar N^s$ on the scaffold occupancy of membrane patches to automate detection of domain boundaries.
We fix the value of $\bar N^s$ by examining how the average receptor and
scaffold numbers per synaptic domain, as well as the average number of synaptic
domains in the system, change as $\bar N^s$ is varied [see Fig.~\ref{fig:patterns0}(b)]. For small enough $\bar N^s$, we obtain a single domain in the system encompassing
all receptors and scaffolds, yielding the global maxima of the average receptor
and scaffold numbers per domain, and the global minimum of the average domain
number. As $\bar N^s$ is increased, the average domain number tends to increase rapidly, because of the many small, transient domains produced by the noise in the system. Concurrently, we find a drop in the average receptor and scaffold numbers per synaptic domain. The average domain number peaks for a value of $\bar N^s$ yielding a maximum number of small, transient domains, at which point we also find local minima in the average receptor and scaffold numbers per synaptic domain. As $\bar N^s$ is increased further, small, transient domains are increasingly filtered out, producing a drop in the average domain number and an increase in the average receptor and scaffold numbers per synaptic domain. We find that the average scaffold number per synaptic domain peaks
at $\bar N^s \approx 0.08$. Beyond this local maximum, the average scaffold number per synaptic domain only decreases gradually with increasing $\bar N^s$. At around $\bar N^s \approx 0.08$ we also find a local maximum in the average receptor number per synaptic domain, as well as a marked decrease in the rate at which the average domain number decreases with increasing $\bar N^s$ [Fig.~\ref{fig:patterns0}(b)]. These results suggest that $\bar N^s = 0.08$ provides a suitable threshold for quantifying the boundaries of the stochastic reaction-diffusion patterns implied by our KMC simulations of the ME~(\ref{eq:ME}), which is also confirmed by direct inspection of our KMC data [Fig.~\ref{fig:patterns0}(a)].

\begin{figure}[t!]
\centering
\includegraphics[width=\columnwidth]{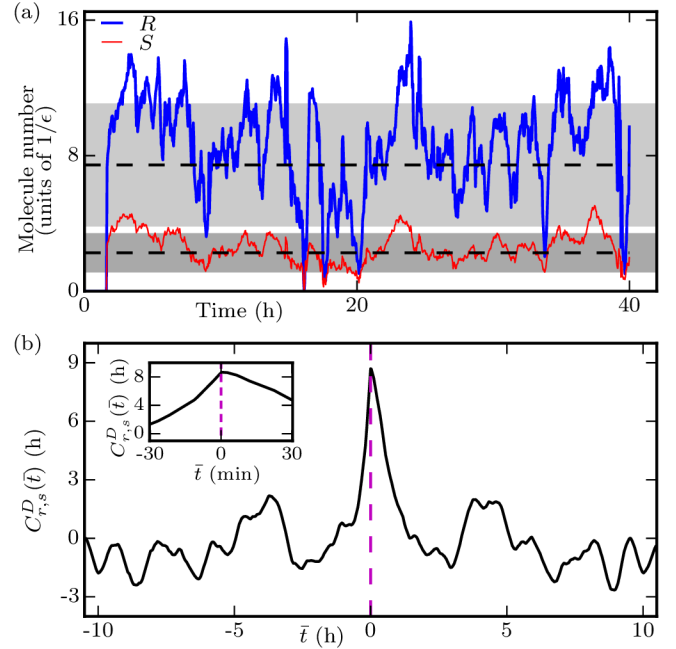} 
\caption{Fluctuating receptor and scaffold numbers in synaptic domains. (a) Receptor and scaffold numbers for the synaptic domain delineated by black domain boundaries in Fig.~\ref{fig:patterns0}(a) versus time using $\bar N^s=0.08$. The horizontal dashed lines and shaded areas indicate the averages and standard deviations of the receptor and scaffold numbers per synaptic
domain, which we obtained from the domains in the upper panels of Fig.~\ref{fig:patterns0}(a). (b) Receptor-scaffold correlation function $C^{D}_{r,s}(\bar{t})$ in Eq.~(\ref{eq:corr_domain}) for the domain delineated by black domain boundaries in Fig.~\ref{fig:patterns0}(a) versus correlation time $\bar t$, using $t_\mathrm{min}=1$~h and $t_\mathrm{max}=40$~h.
The vertical dashed lines in the main panel and inset in panel (b)
correspond to $\bar t=0$.
}
\label{fig:patterns}
\end{figure}

Using $\bar N^s = 0.08$ we find that the mean values and standard deviations
of the receptor and scaffold numbers per synaptic domain are given by $745\pm355$ and $226\pm108$ for the KMC data in Fig.~\ref{fig:patterns0}(a), respectively, with pronounced fluctuations in the in-domain receptor and scaffold population numbers, over a time scale of hours [see Fig.~\ref{fig:patterns}(a)]. In analogy to the receptor-scaffold correlation function defined for the reaction-only system in Eq.~(\ref{eq:correlation}), we quantify the fluctuations in Fig.~\ref{fig:patterns}(a) by calculating the correlation function of the receptor and scaffold populations
in a given synaptic domain,
\begin{align} \label{eq:corr_domain}
C^{D}_{r,s}(\bar{t}) = \int\limits_{t_\mathrm{min}}^{t_\mathrm{max}} dt \left[N^D_r(t)-\langle N^D_r \rangle\right] \left[N^D_s(t+\bar t)-\langle N^D_s \rangle\right] \,,
\end{align}
where the $N^D_{r,s}(t)$ denote the in-domain receptor and scaffold numbers at time $t$. The averages $\langle N^D_{r,s} \rangle$ are evaluated over the time window $t_\mathrm{min}\leq t \leq t_\mathrm{max}$. We choose $t_\mathrm{min}$ and $t_\mathrm{max}$ to correspond to the lifetime of a given synaptic domain at the membrane. In marked contrast to the results for the reaction-only system in Fig.~\ref{fig:full_reac_c}, which yield a time of maximum correlation between fluctuations in the receptor and scaffold occupancies of the order of minutes, we find that the global maximum of $C^{D}_{r,s}(\bar{t})$ occurs at $\left|\bar t\right| < 1$~s [see Fig.~\ref{fig:patterns}(b)]. Thus, our KMC simulations of the ME~(\ref{eq:ME}) suggest that diffusion strongly diminishes the time of maximum correlation between fluctuations in the receptor and scaffold populations at synaptic domains.

\begin{figure}[t!]
\includegraphics[width=\columnwidth]{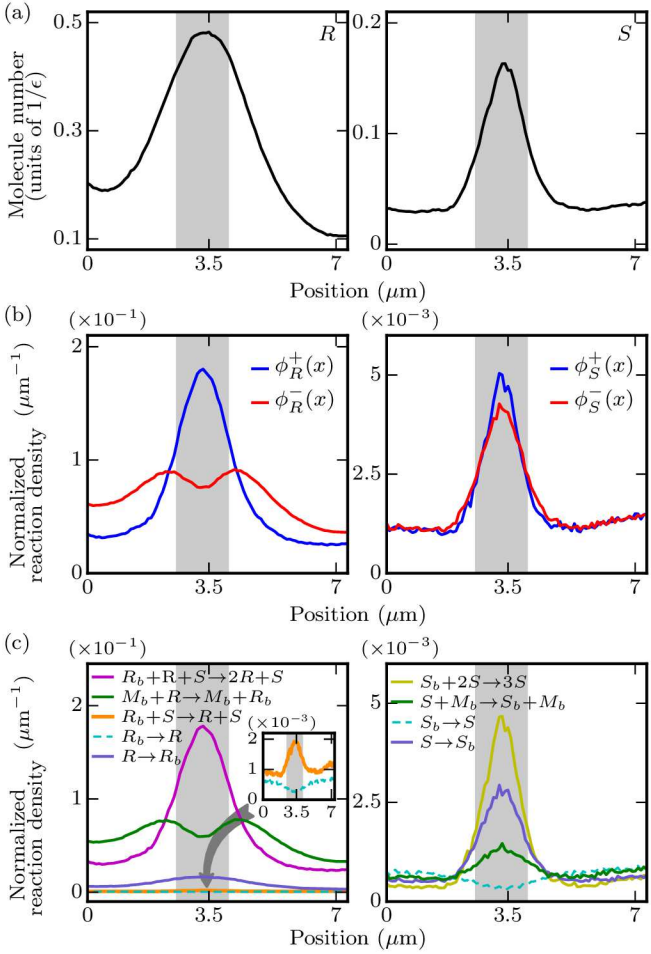} 
\caption{Occurrence probabilities of receptor and scaffold reactions across synaptic domains. (a) Average receptor and scaffold profiles for the synaptic
domain delineated by black domain boundaries in Fig.~\ref{fig:patterns0}(a). (b) Average densities of reactions inserting (removing) receptors and scaffolds into (from) the membrane, $\phi^+_{R,S}$ ($\phi^-_{R,S}$). The average width of the domain is indicated by shaded regions. (c) Average reaction densities
as in panel (b), but for each individual reaction in the reaction-diffusion model considered here (see Sec.~\ref{secModel}). All the results in panels
(b) and (c) were obtained from the domain delineated by black domain boundaries in Fig.~\ref{fig:patterns0}(a), by averaging from $t=1$~h to $t=40$~h in Fig.~\ref{fig:patterns0}(a). The average reaction densities were normalized by the total number of receptor and scaffold reactions that occurred for the spatial region and time interval considered here.}
\label{fig:xcorr}
\end{figure}

We now turn our focus to the occurrence probabilities of receptor and scaffold reactions across synaptic domains. In particular, we consider the
synaptic domain delineated by black boundaries in Fig.~\ref{fig:patterns0}(a),
and first average the corresponding receptor and scaffold profiles over the lifetime of this domain [see Fig.~\ref{fig:xcorr}(a)]. We compute, across the synaptic domain, the densities of all reactions inserting (removing) receptors and scaffolds into (from) the membrane, and average these densities over the lifetime of the domain. We denote the resultant reaction densities by $\phi^+_{R,S}$ ($\phi^-_{R,S}$), which we normalize by the total number of receptor and scaffold reactions that occurred for the spatial region and time interval considered here [see Fig.~\ref{fig:xcorr}(b)]. We find that $\phi^+_{R}$ and $\phi^\pm_{S}$ trace the approximate domain profile with the maxima in $\phi^+_{R}$ and $\phi^\pm_{S}$ occurring close to the domain center, which also shows the largest concentration of receptors and scaffolds [Fig.~\ref{fig:xcorr}(a)]. In contrast, $\phi^-_{R}$ shows a local minimum at the domain center, with local maxima of $\phi^-_{R}$ occurring just outside the synaptic domain. This model prediction is consistent with experimental observations \cite{Blanpied2002} suggesting that receptors are removed from the membrane in membrane regions adjacent to synaptic domains.

To further investigate the origin of the aforementioned qualitative differences in the profiles of $\phi^+_{R}$ and $\phi^\pm_{S}$, and $\phi^-_{R}$, across
synaptic domains, we consider the occurrence probabilities of each individual receptor and scaffold reaction across synaptic domains [see Fig.~\ref{fig:xcorr}(c)]. We find that the two dominant reactions for the receptors are $R_b+R+S\rightarrow2R+S$ and $M_b+R\rightarrow M_b+R_b$, while $S_b+2S\rightarrow3S$ and $S\rightarrow
S_b$ are dominant for the scaffolds. In particular, $\phi^-_{R}$ is mainly set by the reaction $M_b+R\rightarrow M_b+R_b$, which shows a similar profile as $\phi^-_{R}$. Steric constraints reduce the effective rate of the reaction $M_b+R\rightarrow M_b+R_b$ in crowded membrane regions, yielding a local
minimum of the occurrence probability of $M_b+R\rightarrow M_b+R_b$ at the
domain center. However, receptors diffuse rapidly, and can therefore readily
leave synaptic domains. As a result, a substantial number of receptors are
removed via the reaction $M_b+R\rightarrow M_b+R_b$ in the (less crowded) membrane regions adjacent to synaptic domains, producing local maxima of
the occurrence probability of $M_b+R\rightarrow M_b+R_b$ and, hence, $\phi^-_{R}$ close to the domain boundaries. Note that the reaction $M_b+S\rightarrow M_b+S_b$, which is the scaffold reaction analogous to $M_b+R\rightarrow M_b+R_b$,
does not produce pronounced local maxima of $\phi^-_{S}$ close to the domain boundaries. This can be understood by noting that scaffolds diffuse less rapidly than receptors, and are therefore less likely to diffuse into less crowded membrane regions with a large effective rate of $M_b+S\rightarrow M_b+S_b$.

\subsection{Molecular turnover}
\label{secSynMol}

\begin{figure}[b!]
\centering
\includegraphics[width=\columnwidth]{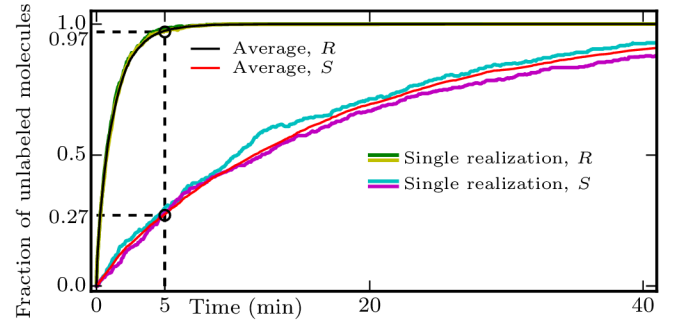} 
\caption{Fractions of unlabeled receptors and scaffolds versus time, obtained from the synaptic domain delineated by black domain boundaries in Fig.~\ref{fig:patterns0}(a) by labeling all the receptors and scaffolds initially localized inside the synaptic domain. Averages were taken over ten realizations corresponding
to different times in Fig.~\ref{fig:patterns0}(a).}
\label{fig:recovery}
\end{figure}

Synaptic domains have been observed 
\cite{Choquet2013,Ziv2014,Kneussel2014,Salvatico2015,Choquet2003,Triller2008,Triller2005,Okabe1999,Gray2006,Calamai2009}
to be in a dynamic steady state, with rapid turnover of individual
receptors and scaffolds. To study, in our stochastic lattice model, receptor and scaffold turnover at synaptic domains we proceed \cite{Kahraman2016} as in FRAP experiments \cite{Choquet2003,Specht2008,Calamai2009}, and label all receptors and scaffolds inside a synaptic domain at a given time. The
fraction of unlabeled receptors and scaffolds inside the synaptic domain
as a function of time, monitored starting from the time when the receptors and scaffolds inside the synaptic domain were labeled, then provides a measure of global molecular turnover at synaptic domains (see Fig.~\ref{fig:recovery}). The time scale in our reaction-diffusion model is set \cite{Haselwandter2011,Haselwandter2015} by the rate of receptor endocytosis. We adjust \cite{Kahraman2016} the rate of receptor endocytosis within the range of values suggested by experiments \cite{Haselwandter2011,Haselwandter2015,Choquet2003,Specht2008,Triller2008} to $k_1=1/750$~$\text{s}^{-1}$ (see Table~\ref{tabReactionRates}) so that,
consistent with FRAP experiments \cite{Choquet2003,Specht2008,Calamai2009},
$\approx30\%$ of scaffolds, but $>95\%$ of receptors, are replaced, on average,
within $5$~min. We find that, on average, $>99\%$ of receptors (scaffolds) are replaced within $\approx7$~min ($\approx 80$~min). Single realizations
of receptor and scaffold turnover at synaptic domains \cite{Kahraman2016}
trace closely, in our model, the corresponding results obtained by averaging over several realizations (Fig.~\ref{fig:recovery}).

\subsection{Single-molecule dynamics}

\begin{figure}[b!]
\centering
\includegraphics[width=\columnwidth]{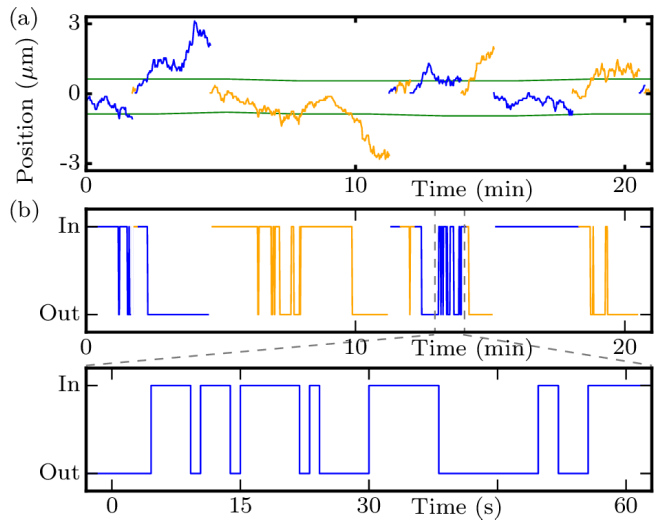} 
\caption{Receptor trafficking along the membrane between the inside and outside
of synaptic domains. (a) Representative trajectories of individual receptors (blue and orange curves) inserted inside the synaptic domain delineated by black domain boundaries in Fig.~\ref{fig:patterns0}(a). Receptors are tracked at the membrane from insertion until removal. Domain boundaries are indicated by green curves. (b) Receptor switching times between the inside (in) and outside (out) of synaptic domains for the diffusion trajectories along the membrane in panel (a).}
\label{fig:indi_traj}
\end{figure}

In agreement with experiments \cite{Choquet2003,Specht2008,Triller2005,Triller2008,Ribrault2011,Choquet2013,Ziv2014,Salvatico2015}, our reaction-diffusion model predicts \cite{Kahraman2016} that receptors initially localized in synaptic domains tend to leave synaptic domains via diffusion, while scaffolds typically stay localized within synaptic domains over their lifetime at the membrane. Following the trajectories of individual
receptors from insertion into the membrane until removal from the membrane,
we find that receptors tend to exchange, via diffusion, between the inside and outside of synaptic domains over their lifetime at the membrane [see Fig.~\ref{fig:indi_traj}(a)]. In particular, our KMC simulations of the ME~(\ref{eq:ME}) imply that receptors traffick along the membrane between the inside and outside of synaptic domains over a time scale of seconds to minutes [see Fig.~\ref{fig:indi_traj}(b)], which is consistent with the switching times between intra- and extra-synaptic
membrane regions typically found in single-molecule experiments on receptor diffusion at synaptic domains 
\cite{Choquet2003,Meier2001,Borgdorff2002,Dahan2003,Triller2005,Specht2008,Triller2008,Choquet2013}.

\section{Summary and conclusions}
\label{secSum}

In neurons, neurotransmitter receptors are concentrated in membrane regions associated with postsynaptic domains \cite{McAllister2007,Ziv2014,Salvatico2015,Citri2008,Legendre2001,Specht2008,Tyagarajan2014}, which are enormously complex molecular assemblies. Experiments on minimal systems devoid of the molecular machinery commonly associated with postsynaptic domains \cite{McAllister2007,Citri2008}---such as single transfected fibroblast cells---have shown \cite{Kirsch1995,Meier2000,Meier2001,Borgdorff2002,Dahan2003,Hanus2006,Ehrensperger2007,Calamai2009,Haselwandter2011}
that receptor-scaffold interactions, together with the diffusion properties of each molecule species at the cell membrane, are sufficient for the formation, stability, and characteristic size of synaptic domains observed
in neurons. The observed self-assembly of synaptic domains of a stable characteristic size can be understood \cite{Haselwandter2011,Haselwandter2015,Turing1952} based on mean-field models of receptor-scaffold reaction-diffusion processes at the cell membrane. However, mean-field models cannot capture how the rapid stochastic dynamics of individual synaptic receptors and scaffolds \cite{Meier2001,Borgdorff2002,Dahan2003,Hanus2006,Triller2005,Specht2008,Triller2008}
relate \cite{Ribrault2011,Choquet2013,Ziv2014,Salvatico2015} to the observed collective properties of synaptic domains. In particular, experiments \cite{Ribrault2011,Choquet2013} and theoretical modeling \cite{Shouval2005,Holcman2006,Sekimoto2009,Burlakov2012,Czondor2012}
suggest that synaptic domains undergo collective fluctuations that may affect synaptic signaling.

In this article we have provided a detailed discussion of a stochastic lattice
model of receptor-scaffold reaction-diffusion processes at the cell membrane \cite{Haselwandter2011,Haselwandter2015}
that yields \cite{Kahraman2016} emergence of synaptic domains in the presence of rapid stochastic turnover of individual molecules. Based on the reaction and diffusion properties of synaptic receptors and scaffolds suggested by previous experiments and mean-field calculations
\cite{Kirsch1995,Meier2000,Meier2001,Borgdorff2002,Dahan2003,Hanus2006,Ehrensperger2007,Calamai2009,Haselwandter2011,Haselwandter2015},
we have shown previously \cite{Kahraman2016} that this stochastic lattice model provides a simple physical mechanism for collective fluctuations in synaptic domains \cite{Ribrault2011,Choquet2013}, the molecular turnover observed at synaptic domains \cite{Choquet2003,Specht2008,Calamai2009}, key features of the observed single-molecule trajectories 
\cite{Choquet2003,Choquet2013,Meier2001,Borgdorff2002,Dahan2003,Hanus2006,Specht2008,Triller2005,Triller2008}, and spatially inhomogeneous receptor and scaffold lifetimes at the cell membrane \cite{Czondor2012,Blanpied2002,Earnshaw2006}. We have confirmed here these conclusions \cite{Kahraman2016}, and expanded and elaborated upon our previous results, using a combination of KMC simulations of our stochastic lattice model of receptor-scaffold reaction-diffusion processes at synaptic domains,
and analytic and numerical solutions of the ME governing our stochastic lattice model.

For the diffusion-only system we find that, even for $1/\epsilon=8$, which
corresponds to a maximum molecule occupancy per lattice site of only eight
molecules, the average results of KMC simulations are in quantitative agreement with the corresponding mean-field model \cite{Haselwandter2011,Haselwandter2015,Satulovsky1996,McKane2004,Lugo2008,Fanelli2010,Fanelli2013} of nonlinear diffusion in crowded membranes. A possible origin for this agreement between the stochastic lattice model and the mean-field model, even for large $\epsilon$, is that, in the diffusion-only system, the molecule number is conserved, which constrains the fluctuations in the stochastic system \cite{Haselwandter2002}. The ME~(\ref{eq:ME}) with $W_\textrm{react}=0$ and the corresponding mean-field equations~(\ref{eq:MFE_r}) and~(\ref{eq:MFE_s}) with $F^r=F^s=0$ 
\cite{Haselwandter2011,Haselwandter2015,Satulovsky1996,McKane2004,Lugo2008,Fanelli2010,Fanelli2013}
produce non-Gaussian and, in some cases, even non-monotonic diffusion profiles. In particular, compared to Fickian diffusion, crowding tends to yield less disperse molecule distributions \cite{Kahraman2016}, with the slowly-diffusing
scaffolds acting as an effective barrier to the dispersal of the more rapidly diffusing receptors \cite{Meier2001,Borgdorff2002,Dahan2003,Hanus2006,Triller2005,Specht2008,Triller2008,Choquet2013,Ziv2014,Salvatico2015}.
Calculation of the MSD of individual receptors and scaffolds shows that,
at finite times, the MSD bears a signature of the initial distribution of
the receptors and scaffolds in the system. As the system approaches its steady
state, with a homogeneous distribution of receptors and scaffolds, we recover
the characteristics of the MSD associated with standard Fickian diffusion, provided that the diffusion
coefficient is rescaled by a factor accounting for the average molecular crowding in the system.

As discussed in Sec.~\ref{secModel}, for synaptic domains formed by glycine receptors and gephyrin, a physically reasonable choice for the value of $\epsilon$ is $\epsilon\approx1/100$ \cite{Kahraman2016} so that, for a size $a_P \approx5$--$10$~nm of glycine receptors and gephyrin \cite{Kim2006,Du2015}, the membrane patch size $a\approx 80$~nm is smaller than the expected typical size of synaptic domains 
\cite{Kirsch1995,Meier2000,Meier2001,Borgdorff2002,Dahan2003,Hanus2006,Ehrensperger2007,Calamai2009,Haselwandter2011,Haselwandter2015}.
As $\epsilon \to 0$, one generally expects that averages over the ME~(\ref{eq:ME}) coincide with the solutions of the corresponding mean-field equations~(\ref{eq:MFE_r}) and~(\ref{eq:MFE_s}) \cite{Haselwandter2011,Haselwandter2015}. As summarized
above, agreement between stochastic lattice model and mean-field model is already obtained, for the diffusion-only system, with $\epsilon \approx 1/10$. In contrast, for the reaction-only system we find that, for physically relevant values of $\epsilon$, averages over the stochastic lattice model do, in general, not coincide with solutions of the corresponding mean-field equations. Exceptions to this conclusion are provided by purely linear reaction schemes with only a single linear reaction or competing linear reactions yielding a fluctuating steady state of the system (and no absorbing state). Indeed, our exact analytic solutions of the ME show that, in the case of competing linear reactions with a fluctuating steady state, agreement between ME and mean-field equations can even be obtained with $\epsilon=1$, which corresponds to a maximum molecule occupancy per membrane patch of only
one molecule.

Consistent with previous theoretical studies of stochastic reaction-diffusion systems
\cite{Samoilov2006,Erban2009,Gillespie1976,Gillespie1977,Gillespie2013,McKane2004,Lugo2008,Kahraman2016},
we find that molecular noise generally plays a central role in the reaction dynamics at synaptic domains. In particular, for the nonlinear reactions thought to be relevant for synaptic domains (see Sec.~\ref{secModel}) we
find, for physically relevant values of $\epsilon$, disagreement between solutions of the ME~(\ref{eq:ME}) with $W_\textrm{diff}=0$
and the corresponding mean-field equations~(\ref{eq:MFE_r}) and~(\ref{eq:MFE_s}) with $\nu_r=\nu_s=0$ \cite{Haselwandter2011,Haselwandter2015,Satulovsky1996,McKane2004,Lugo2008,Fanelli2010,Fanelli2013}.
For instance, trimerization of scaffolds, $S_b+2S\rightarrow3S$, is found \cite{Choquet2003,Calamai2009,Kahraman2016,Haselwandter2011,Haselwandter2015}
to be crucial for the self-assembly of stable synaptic domains composed of
glycine receptors and gephyrin. Even in a very simple reaction scheme, in
which only this single reaction is considered, a value $\epsilon \lessapprox 1/5000$ is required to produce reasonable agreement between the stochastic
model and the mean-field model. Such a value of $\epsilon$ implies a membrane patch size $a\approx 71 a_P \approx 350$--$710$~nm for glycine receptors and gephyrin \cite{Kim2006,Du2015}. The assumption of a well-mixed system
is not expected to be warranted over such large membrane patch sizes, resulting in breakdown of the mean-field approach \cite{Epstein1998,Walgraef1997,Cross2009,Cross1993,Murray2002,Meinhardt1982,Maini2001}
even for single (nonlinear) chemical reactions. In addition to the amplification of noise through nonlinear chemical reactions with steric constraints \cite{Kahraman2016,Erban2009}, illustrated by $S_b+2S\rightarrow3S$, we also \cite{Samoilov2006} find that absorbing (non-fluctuating) states and bistability provide physical mechanisms yielding disagreement between stochastic and mean-field models of the reaction dynamics at synaptic domains. In the case of competing absorbing states we find that the mean-field model can produce quantitative agreement with averages over the stochastic model while failing to reproduce the asymptotic properties of individual realizations of the stochastic system, with distinct stochastic trajectories of the system being trapped in distinct absorbing states.

Comparing stochastic and mean-field results for the complete reaction dynamics at synaptic domains (see Sec.~\ref{secModel}) we find \cite{Kahraman2016} that, for $\epsilon \approx 1/100$, the mean-field equations~(\ref{eq:MFE_r}) and~(\ref{eq:MFE_s}) with $\nu_r=\nu_s=0$ fail to capture the temporal evolution as well as steady-state values of the average receptor and scaffold occupancies implied by the ME~(\ref{eq:ME}) with $W_\textrm{diff}=0$, with the average stochastic dynamics being approximately
one order of magnitude faster than the mean-field dynamics. Calculation of the marginal steady-state probability distribution of the receptor occupancy
shows that one origin for this discrepancy between stochastic and mean-field solutions lies, for $\epsilon \gtrapprox 1/300$, in the bistability of the stochastic system. For $\epsilon \lessapprox 1/300$, we find a single mode of the marginal steady-state probability distribution of the receptor occupancy. However, our KMC simulations show that, even for $\epsilon \approx 7.8\times10^{-6}$, the averages as well as the modes of the receptor and scaffold occupancies can disagree substantially with mean-field results. We find that, in our stochastic model, the receptor and scaffold occupancies can both undergo large fluctuations, with the fluctuations in the receptor occupancy being particularly pronounced. Calculation of the receptor-scaffold correlation
function shows that, consistent with the roles of receptors and scaffolds
as inhibitors and activators of increased molecule concentrations \cite{Haselwandter2011,Haselwandter2015},
fluctuations increasing (decreasing) the scaffold occupancy tend to lead to an increase (a decrease) in the receptor occupancy, with the time of maximum correlation between receptor and scaffold fluctuations being of the order of minutes. We find that the time of maximum correlation between receptor and scaffold fluctuations tends to increase with decreasing $\epsilon$.

Finally, we considered in this article the receptor-scaffold reaction-diffusion dynamics at synaptic domains suggested by previous experiments and mean-field calculations 
\cite{Kirsch1995,Meier2000,Meier2001,Borgdorff2002,Dahan2003,Hanus2006,Ehrensperger2007,Calamai2009,Haselwandter2011,Haselwandter2015}. We find that, starting from random initial conditions, the stochastic lattice model described by the ME~(\ref{eq:ME})
yields \cite{Kahraman2016} spontaneous formation of in-phase receptor and
scaffold domains with a characteristic wavelength consistent with the corresponding mean-field results implied by Eqs.~(\ref{eq:MFE_r}) and~(\ref{eq:MFE_s}).
We have shown previously \cite{Haselwandter2011,Haselwandter2015,Kahraman2016}
that, in 2D, the mean-field equations~(\ref{eq:MFE_r}) and~(\ref{eq:MFE_s}) and the corresponding ME~(\ref{eq:ME}) yield, for the reaction and diffusion rates suggested by experiments \cite{Kirsch1995,Meier2000,Meier2001,Borgdorff2002,Dahan2003,Hanus2006,Ehrensperger2007,Calamai2009,Haselwandter2011,Haselwandter2015}
and used here (see Sec.~\ref{secModel}),
synaptic domains of a similar characteristic size as observed in experiments
on neurons and transfected fibroblast cells 
\cite{Kirsch1995,Meier2000,Meier2001,Borgdorff2002,Dahan2003,Hanus2006,Ehrensperger2007,Calamai2009,Haselwandter2011}.
Consistent with our results for the reaction-only system, we find \cite{Kahraman2016},
for $\epsilon\approx1/100$, that molecular noise accelerates synaptic domain formation by approximately one order of magnitude compared to mean-field dynamics while producing,
over a time scale of several hours, substantial fluctuations in the size and location of synaptic domains. These results illustrate the potential importance of stochastic effects \cite{Choquet2013} when describing synaptic
domains. We also find that, consistent with experimental observations \cite{Ribrault2011,Choquet2013}, the molecular noise induced by the underlying reaction and diffusion dynamics of synaptic receptors and scaffolds can produce \cite{Kahraman2016} collective fluctuations in synaptic domains. Calculation of the correlation function
of the in-domain receptor and scaffold population numbers shows that, in contrast to our results for the reaction-only system, the reaction-diffusion
system only yields a very short time of maximum correlation between receptor and scaffold fluctuations. Thus, our results suggest that diffusion strongly diminishes the correlation time between fluctuations in the receptor and scaffold populations at synaptic domains~\cite{Ribrault2011,Choquet2013}.

We find \cite{Kahraman2016} that, in both our mean-field and stochastic models, scaffold profiles across synaptic domains tend to be more narrow than receptor profiles. Scaffold domains therefore tend to be more sharply defined than receptor domains, and we quantify \cite{Kahraman2016} domain boundaries in our stochastic lattice model by placing a threshold on the scaffold occupancy per membrane patch. Our stochastic lattice model allows us to compute the occurrence probabilities of receptor and scaffold reactions across synaptic domains. We find that, even though the reaction rates in our reaction-diffusion model are constant, the occurrence probabilities of the receptor-scaffold
reactions considered here can be strongly inhomogeneous across synaptic domains. In particular, we find that the occurrence probabilities
of reactions decreasing/increasing the scaffold number, and increasing the
receptor number, trace the approximate domain profile, with maxima close to the domain center. In contrast the occurrence probability of reactions decreasing the receptor number is minimal at the domain center, and shows
local maxima just outside the synaptic domain. Thus, our model predicts that
receptors tend to be removed from the cell membrane in membrane regions adjacent to synaptic domains, which is consistent with experimental observations \cite{Blanpied2002}.

As discussed in Sec.~\ref{secRSD}, our stochastic lattice model provides a simple explanation for spatial heterogeneity in the occurrence probabilities of receptor and scaffold reactions across synaptic domains \cite{Czondor2012,Blanpied2002,Earnshaw2006} in terms of steric constraints on the receptor and scaffold membrane patch occupancies, together with the reaction-diffusion instability \cite{Haselwandter2011,Haselwandter2015,Turing1952} of the model discussed here. Similarly, we have shown previously \cite{Kahraman2016}
that our stochastic lattice model provides a simple physical mechanism for distinct receptor and scaffold lifetimes at the membrane inside and outside synaptic domains \cite{Czondor2012,Blanpied2002,Earnshaw2006}. We also find \cite{Kahraman2016} that, based on the reaction and diffusion properties of synaptic receptors and scaffolds suggested by previous experiments and mean-field calculations 
\cite{Kirsch1995,Meier2000,Meier2001,Borgdorff2002,Dahan2003,Hanus2006,Ehrensperger2007,Calamai2009,Haselwandter2011,Haselwandter2015},
the stochastic lattice model discussed here can yield the turnover times
of receptor and scaffold populations observed at synaptic domains \cite{Choquet2003,Specht2008,Calamai2009},
and predicts \cite{Kahraman2016} single-molecule trajectories consistent with experimental observations
\cite{Choquet2003,Choquet2013,Meier2001,Borgdorff2002,Dahan2003,Specht2008,Triller2005,Triller2008}.

Many essential cellular processes rely on the organization of membrane proteins into membrane protein domains \cite{Lang2010,Simons2010,Rao2014,Recouvreux2016}.
Membrane protein domains are characterized
\cite{Lang2010,Simons2010,Rao2014,Recouvreux2016,Choquet2003,Specht2008,Triller2005,Triller2008,Ribrault2011,Choquet2013,Ziv2014,Salvatico2015}
by low protein copy numbers ($\approx10$--1000) and protein crowding. Using synaptic membrane protein domains \cite{Ziv2014,Salvatico2015} as a model system, we have studied here in detail a stochastic lattice model \cite{Kahraman2016} of protein reaction-diffusion processes in crowded cell membranes. Our stochastic lattice model links the molecular noise inherent in reaction-diffusion processes to collective fluctuations in synaptic domains, and allows prediction of the stochastic dynamics of individual synaptic receptors and scaffolds.
We find that molecular noise can yield substantial deviations from mean-field results, and that stochastic lattice models can be employed successfully to provide quantitative insights into single-molecule and collective dynamics of membrane protein domains \cite{Lang2010,Simons2010,Rao2014,Recouvreux2016,Choquet2003}.
We focused here on the most straightforward scenario of a 1D system, which already captures \cite{Kahraman2016} the basic phenomenology of the observed fluctuations at synaptic domains. Generalization of our KMC simulations to 2D systems, which can be handled efficiently \cite{Kahraman2016} with the computational approach \cite{Elf2003} we use here, will allow more detailed and quantitative model predictions pertaining to, for instance, the stochastic trajectories of individual synaptic receptors and scaffolds at the cell membrane, which can now be directly measured \cite{Hell1994,Choquet2003,Huang2010,Kusumi2014} in experiments
\cite{Choquet2003,Choquet2013,Meier2001,Borgdorff2002,Dahan2003,Hanus2006,Specht2008,Triller2005,Triller2008}.
Our work sheds light on the organizational principles linking the collective properties of biologically important supramolecular structures, such as synaptic membrane protein domains, to the stochastic dynamics that rule their molecular components.

\acknowledgments{We thank R. A. da Silveira, M. Kardar, and A. Triller for helpful discussions. This work was supported by NSF award numbers DMR-1554716 and DMR-1206332, an Alfred P. Sloan Research Fellowship in Physics, the James H. Zumberge Faculty Research and Innovation Fund at USC, and the USC Center for High-Performance Computing. We also acknowledge support through the Kavli Institute for Theoretical Physics, Santa Barbara, via NSF award number PHY-1125915.}

\appendix

\section{Probability distribution of jump times in a chain of Poisson processes}
\label{Appendix A}

For completeness, we compute here the probability distribution of jump times
between arbitrary states in the chain of Poisson processes in Eq.~(\ref{eq:Markov_chain}).
We first note that
\begin{align} \label{eq:convolution}
 P_{i\rightarrow i+2}(t) = \int dt' P_{i\rightarrow i+1}(t') P_{i+1\rightarrow i+2}(t-t') \, .
\end{align}
Taking the Laplace transform
\begin{equation}
\tilde{P}_{i\to j}(s)\equiv\mathcal{L}\left[P_{i\to j}(t)\right](s)=\int\limits_0^\infty dt e^{-st}P_{i\to j}(t)\,,
\end{equation}
the convolution in Eq.~(\ref{eq:convolution}) can be conveniently represented in the form
\begin{align} \label{eq:convolution2}
 \tilde{P}_{i\rightarrow i+2}(s) = \tilde{P}_{i\rightarrow i+1}(s) 
          \times \tilde{P}_{i+1\rightarrow i+2}(s) \, .
\end{align}
From Eq.~(\ref{eq:EXP_dist}) we have
\begin{align}
 \tilde{P}_{i\rightarrow i+1}(s) = \int\limits_0^\infty \alpha_i e^{-(\alpha_i+s)t} dt = \frac{\alpha_i}{\alpha_i+s} \,,
\end{align}
which implies that
\begin{align} \label{eq:Laplace}
\tilde{P}_{i\rightarrow i+2}(s) = \frac{\alpha_i}{\alpha_i+s} \frac{\alpha_{i+1}}{\alpha_{i+1}+s} \, .
\end{align}
Transforming Eq. (\ref{eq:Laplace}) back by using an inverse Laplace transform,
one obtains
\begin{align}
P_{i\rightarrow i+2}(t) &= \mathcal{L}^{-1}\left[\tilde{P}_{i\rightarrow i+2}(s)\right](t) \nonumber\\
   &= \frac{\alpha_i \alpha_{i+1}}{\alpha_{i+1}-\alpha_{i}} \left(e^{-\alpha_it} - e^{-\alpha_{i+1}t}\right) \, .
\end{align}
Following similar steps as outlined above, the probability distribution of jump times for an $n$-step jump can be expressed as an $n$-fold convolution, resulting in
\begin{align}
 P_{i\rightarrow i+n}(t) =& \mathcal{L}^{-1}\left[\prod_{j=i}^{i+n-1} \frac{\alpha_j}{\alpha_j+s} \right](t) \nonumber \\
 =& \left(\prod_i \alpha_i \right) \sum\limits_i \frac{e^{-\alpha_i t}}{\prod_{j\neq i} \left(\alpha_j-\alpha_i\right)} \, ,
\label{eq:sol_sing_reac}
\end{align}
which provides the probability distribution of jump times between arbitrary states in the chain of Poisson processes in Eq.~(\ref{eq:Markov_chain}).

\section{Mean jump time in a chain of Poisson processes}
\label{Appendix B}

In this appendix we formally derive the expression in Eq.~(\ref{eq:avg_jump_sol2MT})
for the average jump time from a state $p$ to a state $q$, $\langle t \rangle_{p\to q}$, in the chain of Poisson processes in Eq.~(\ref{eq:Markov_chain}). Consider the mean jump time from state $i$ to state $i+2$ in Eq.~(\ref{eq:Markov_chain}):
\begin{align} \label{eq:meantimeAppendix}
 \langle t \rangle_{i\to i+2} &= \int\limits_0^\infty dt \, t P_{i\to i+2}(t)  \nonumber\\ 
 &= \int\limits_0^\infty dt \, t  \int\limits_0^t dt' P_{i\to i+1}(t') P_{i+1\to i+2}(t-t')  \nonumber\\ 
 &= \int\limits_0^\infty dt' P_{i\to i+1}(t')  \int\limits_{t'}^\infty dt
 \, t
 P_{i+1\to i+2}(t-t')  \nonumber \\
 &= \int\limits_0^\infty dt' P_{i\to i+1}(t') \int\limits_0^\infty dt \left(t'+t\right)P_{i+1\to i+2}(t) \nonumber \\
 &= \int\limits_0^\infty dt' \left(t'+\langle t \rangle_{i+1\to i+2}\right)P_{i\to i+1}(t')  \nonumber \\
 &= \langle t \rangle_{i,i+1} + \langle t \rangle_{i+1,i+2} \,,
\end{align}
where we used that $\int_0^\infty dt P_{j\to j+1}(t) = 1$. Decomposing an
arbitrarily long sequence of jump processes into pairs of jump processes and using Eq.~(\ref{eq:meantimeAppendix}), Eq.~(\ref{eq:avg_jump_sol2MT}) follows from Eq.~(\ref{eq:EXP_dist}). Using Eq.~(\ref{eq:sol_sing_reac}), Eq.~(\ref{eq:avg_jump_sol2MT}) can also be derived directly from the probability distribution of jump times for an $n$-step jump in Eq.~(\ref{eq:Markov_chain}).


\end{document}